%% file: ms.tex
\documentclass[journal, preprint]{vgtc}                 

\ifpdf
  \pdfoutput=1\relax                   
  \usepackage{graphicx}                
  \DeclareGraphicsExtensions{.pdf,.png,.jpg,.jpeg} 
\else
  \usepackage{graphicx}                
  \DeclareGraphicsExtensions{.eps}     
\fi%

\graphicspath{{figures/}}              

\PassOptionsToPackage{warn}{textcomp}  
\usepackage{textcomp}                  
\usepackage{mathptmx}                  
\usepackage{times}                     
\usepackage{cite}                      
\usepackage{tabu}                      
\usepackage{booktabs}                  

\usepackage{amsfonts}
\usepackage{amsmath}
\usepackage{algorithm}
\usepackage{algorithmic}
\usepackage{caption}
\usepackage{xspace}

\input{macros}

\onlineid{0}

\vgtccategory{Research}

\input{title}

\input{abstract}

\C{
\CCScatlist{
  \CCScatTwelve{Error-bounded lossy data compression}{Progressive data retrieval}
}
}

\begin{document}


\input{intro}
\input{related}
\input{preliminaries}
\input{method}
\input{results}
\input{conclusion}
\input{acknowledgments}
\clearpage


\bibliographystyle{abbrv-doi-hyperref}

\bibliography{ms}
\end{document}

%% file: macros.tex
\newcommand{\C}[1]{} 

\newcommand{\codename}[1]{{\textsc{#1}}\xspace}
\newcommand{\zfp}{\codename{zfp}}
\newcommand{\zfpr}{\codename{zfp-r}}
\newcommand{\mzfp}{\codename{mzfp}}
\newcommand{\sperr}{\codename{sperr}}
\newcommand{\msperr}{\codename{msperr}}
\newcommand{\sz}{\codename{sz}}
\newcommand{\msz}{\codename{msz}}
\newcommand{\szthree}{\codename{sz3}}
\newcommand{\mgard}{\codename{mgard}}
\newcommand{\mmgard}{\codename{mmgard}}
\newcommand{\pmgard}{\codename{pmgard}}
\newcommand{\mdr}{\codename{mdr}}
\newcommand{\hire}{\codename{hire}}
\newcommand{\tthresh}{\codename{tthresh}}
\newcommand{\fpzip}{\codename{fpzip}}
\newcommand{\vapor}{\codename{vapor}}
\newcommand{\idxtwo}{\codename{idx2}}
\newcommand{\spiht}{\codename{spiht}}
\newcommand{\speck}{\codename{speck}}
\newcommand{\ezw}{\codename{ezw}}
\newcommand{\posits}{\codename{posits}}
\newcommand{\proj}{\codename{proj}}

\renewcommand{\vec}[1]{\mathbf{#1}}
\newcommand{\func}[1]{\mathbb{#1}}

\newcommand{\fl}[1][]{
  \def\@tmp{#1}%
  \ifx\@tmp\empty%
    \ensuremath{\mathrm{fl}}%
  \else%
    \ensuremath{\mathrm{fl}(#1)}%
  \fi%
  \xspace%
}
\newcommand{\error}[1][]{
  \def\@tmp{#1}%
  \ifx\@tmp\empty%
    \ensuremath{e}%
  \else%
    \ensuremath{e_{#1}}%
  \fi%
  \xspace%
}
\newcommand{\comp}[1]{\ensuremath{x_{#1}}\xspace}
\newcommand{\csum}[1]{\ensuremath{\tilde{x}_{#1}}\xspace}
\newcommand{\actual}{\ensuremath{x}\xspace}

\newcommand{\verror}[1][]{
  \def\@tmp{#1}%
  \ifx\@tmp\empty%
    \ensuremath{\mathbf{e}}%
  \else%
    \ensuremath{\mathbf{e}_{#1}}%
  \fi%
  \xspace%
}
\newcommand{\vcomp}[1]{\ensuremath{\mathbf{x}_{#1}}\xspace}
\newcommand{\vzcomp}[1]{\ensuremath{\mathbf{y}_{#1}}\xspace}
\newcommand{\vcsum}[1]{\ensuremath{\tilde{\mathbf{x}}_{#1}}\xspace}
\newcommand{\vactual}{\ensuremath{\mathbf{x}}\xspace}
\newcommand{\vapprox}{\ensuremath{\tilde{\mathbf{x}}}\xspace}
\newcommand{\tol}{\ensuremath{\tau}\xspace}

\newcommand{\linf}[1]{\ensuremath{\|#1\|_\infty}\xspace}

\newcommand{\gain}{\ensuremath{\alpha}\xspace}
\newcommand{\rmse}{\ensuremath{E}\xspace}
\newcommand{\rate}{\ensuremath{R}\xspace}
\newcommand{\gran}{\ensuremath{\Delta}\xspace}
\newcommand{\snr}{\ensuremath{\mathit{SNR}}\xspace}

\newcommand{\github}[1]{#1}

\providecommand{\etal}{et al.\@\xspace}

%% file: title.tex
\title{A General Framework for \linebreak Progressive Data Compression and Retrieval}

\author{%
  \authororcid{Victor A. P. Magri}{0000-0002-3389-523X} and
  \authororcid{Peter Lindstrom}{0000-0003-3817-4199}
}

\authorfooter{%
  \item Victor Paludetto Magri and Peter Lindstrom are with
        Lawrence Livermore National Laboratory.
        E-mail: \{paludettomag1\,$|$\,pl\}@llnl.gov.
}

\teaser{%
  \centering%
  \includegraphics[width=\linewidth]{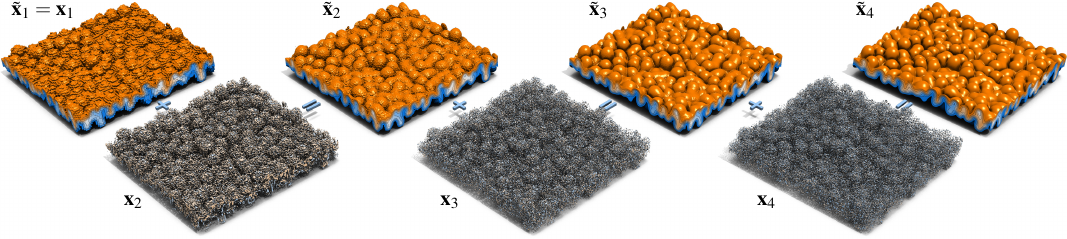}%
  \caption{%
    Multi-component expansion $(\vcomp{1}, \vcomp{2}, \vcomp{3}, \vcomp{4})$
    of a 3D density field.  Progressive reconstructions $\vcsum{i} = \vcsum{i-1} + \vcomp{i}$,
    with $\vcsum{0} = \vec{0}$, are formed by adding components, \vcomp{i}, with
    exponentially decreasing norm.  Components are compressed independently, here to bit rates
    $\{0.40, 0.13, 0.23, 0.39\}$ bits/value (from left to right), using a general-purpose
    lossy compressor, here \zfp~\cite{Li14}.
  }
  \label{fig:teaser}
}

%% file: abstract.tex
\abstract{In scientific simulations, observations, and experiments, the cost of
transferring data to and from disk and across networks has become a significant bottleneck
that particularly impacts subsequent data analysis and visualization. To address this
challenge, compression techniques have been widely adopted. However, traditional lossy
compression approaches often require setting error tolerances conservatively to respect
the numerical sensitivities of a wide variety of post hoc data analyses, some of which may
not even be known a priori. Progressive data compression and retrieval has emerged as a
solution, allowing for the adaptive handling of compressed data according to the needs of
a given post-processing task. However, few analysis algorithms natively support
progressive data processing, and adapting compression techniques, file formats,
client/server frameworks, and APIs to support progressivity can be challenging.
This work presents a general framework that supports progressive-precision data queries
independently of the underlying data compressor or number representation. Our approach is
based on a multiple-component representation that successively, with each new component,
reduces the error between the original and compressed field, allowing each field in the
progressive sequence to be expressed as a partial sum of components. We have implemented
our approach on top of four popular scientific data compressors and have evaluated its
behavior on several real-world data sets from the SDRBench collection. Numerical results
indicate that our framework is effective in terms of accuracy compared to each of the
standalone compressors it builds upon. In addition, compression and decompression time are
proportional to the number and granularity of components, as requested by the
user. Finally, our framework allows for fully lossless compression using lossy compressors
when a sufficient number of components are employed.}

%% file: intro.tex
\firstsection{Introduction}

\maketitle

With the arrival of the exascale era in high-performance computing, achieving efficient
hardware utilization is no longer solely dependent on compute power. Instead, one needs to
account for the cost of data movement between various components of a computer system,
such as DRAM and registers, CPU cores and accelerators, distributed compute nodes, and
main memory and offline storage. It is not practical to record every byte produced in
scientific observations, experiments, and simulations, so data must be reduced using
compression methods that eliminate redundancy and bits of marginal accuracy in
floating-point arrays.

To address this challenge, numerous lossy compressors for floating-point data have been
developed in recent years,
including \mgard~\cite{mgard}, \sperr~\cite{sperr}, \sz~\cite{LiZhSh23}, and
\zfp~\cite{Li14}, to name a few. These compressors are typically inserted at select points
in the memory hierarchy, usually near offline storage, and compression parameters are set
for a specific use case, such as checkpointing, data analysis, or visualization. In this
context, decompressing data for subsequent processing generally requires the transfer and
decompression of whole fields, which incurs expensive data movement and (at least
temporary) storage costs that compression techniques were designed to
alleviate. Furthermore, the output of a single simulation, observation, or experiment may
feed into multiple analyses with varying requirements in resolution, precision, and
regions of interest in space and time.

The data thus obtained at great expense is often shared across facilities, e.g., via
community databases, to benefit the community at large, allowing both intended and
unforeseen scientific studies to be carried out---a few example domains include
climate~\cite{lens,cmip6}, astronomy~\cite{lsst}, particle physics~\cite{cern}, and
turbulence~\cite{jhtdb}.  Faced with a diversity of application domains and use cases with
varying requirements on data fidelity (e.g., precision and resolution), lossy
compression---when integrated---is often used conservatively to cater to the least common
denominator of known or anticipated accuracy needs.  This ``one-size-fits-all'' approach
can waste resources, especially for downstream data analysis applications that do not
require or even benefit from extraneous precision or resolution, yet must query and
transfer full-fidelity data from a central repository.

One proposed solution to combat the high cost of data transfer is to use
\emph{progressive compression} that allows extracting reduced-fidelity data with a
smaller compressed storage and transfer
cost~\cite{Clyne2012,LiGoCh21,jpeg2000,Hoang2019,amm}.  This allows the data to be
archived at or near full fidelity, with each client (data consumer) specifying its
fidelity needs so that only as much compressed data as necessary is transferred.  Usually,
a progressive representation also allows incrementally refining an approximation of the
data as more compressed data arrives.  Unfortunately, while ubiquitous for image
transmission over the web, few scientific databases make use of progressive compression
for several reasons:
\begin{itemize}
   \item Popular file formats and I/O libraries like HDF5~\cite{hdf5}, NetCDF~\cite{netcdf},
ADIOS~\cite{adios2}, and ROOT~\cite{root}, while supporting compression, do not come with
progressive-access support.  Hence, one must forego such tools and instead rely on custom
file formats and I/O libraries.

   \item Few compressors intrinsically support progressive access, with \mgard~\cite{mgard}
and \vapor~\cite{Li2019} being notable exceptions rather than the norm.

   \item Even though compressors based on wavelets (for progressive resolution) or bit plane
coding (for progressive precision) could be made to support progressive access, the
engineering and implementation challenges can be substantial.  Moreover, the
(de)compressor API and implementation must be modified to support such queries, and
usually a specialized client/server layer is needed to coordinate the transfer and
integration of data that will ultimately be served to the science application requesting
it.
\end{itemize}

In this paper, we advocate for a much simpler but pragmatic approach to
progressive-precision queries that supports \emph{any} lossy compressor or number
representation without the need for invasive code changes or custom file format.  Our
framework is based on the idea of \emph{multi-component} decomposition---well-known in
fields like multiple-precision arithmetic, with \emph{double-double}~\cite{Hida:2001}
being perhaps the best-known representation---but generalizes the use of a vector of
low-precision floating-point numbers to represent a high-precision scalar by building on
top of lossy numerical compression.  The salient aspects of our framework are:
\begin{itemize}
   \item A multi-dimensional scalar field is expressed as a sum of components with
(typically) exponentially increasing accuracy.  Each component represents the remaining
error from the partial sum of components preceding it, and is compressed independently by
a lossy compressor of the user's choosing.

   \item We require no special file format, API, or client/server framework.  Rather,
the components constitute an additional data dimension, with the convention that the
consumer requests a subset of compressed components that are decompressed and added to
form an approximation.  Existing compression plugins for formats like HDF5 can thus be
reused as is.

   \item Incremental refinement is supported trivially by simply ``adding in''
components to the reconstruction as they arrive.  Hence, no additional decompressor state
or previously received compressed data needs to be buffered.

   \item Our approach supports error tolerances, with the application requesting
only as many components as needed to satisfy its accuracy requirements. In the limit, we
can ensure lossless compression via lossy compressors if desired.

   \item The granularity of progressivity can be chosen by both the data producer
and, later, by the data consumer in terms of the total number of components stored or
number of components consumed together.  Usually, there is a tradeoff between granularity
supported and overhead in storage and compute time.
\end{itemize}

We detail our approach after reviewing related work on progressive compression of
numerical data.

%% file: related.tex
\section{Related Work}
\label{sec:related}

In this section, we review related work on progressive compression in the context of
scientific data.  In particular, we limit the discussion to multidimensional scalar fields
defined on structured grids.  For a more general discussion of lossy compression and
unstructured data, we refer the reader to the recent survey of Li~\etal~\cite{Li2018}.

Progressive representations allow the reconstruction of a field by decompressing only a
subset of the compressed bit stream and by incrementally refining the reconstruction as
additional compressed bits are processed. Usually, the bit stream is organized such that
early bits have the most impact on accuracy and such that any prefix of the bit stream
gives near-optimal quality at that given bit rate.  There are currently two dominant
approaches: progression in \emph{resolution} (in space, time, or both), usually with
low-frequency components transmitted first, and progression in \emph{precision}, where the
most significant bits are processed first.

In the former camp, numerous multiresolution hierarchies have been proposed, including
those based on subsampling~\cite{idx}, spectral transformations like the Fourier and
discrete cosine transform~\cite{dctz}, Laplacian pyramids~\cite{Burt1987}, wavelets and
related sub-band decompositions~\cite{Li2019,sperr,mgard}, as well as rank decompositions
like SVD and HOSVD~\cite{tthresh,TuckerMPI}, which tend to generate basis functions that
loosely correspond to frequency.  Oftentimes such representations support progression by
simply thresholding or truncating basis coefficients, essentially performing a form of
low-pass filtering, with subsequent coefficients introducing higher-frequency details.

Whereas in the first camp, some subset of coefficients are encoded at full precision,
approaches in the second camp are typified by encoding some subset of ``bit planes'' (the
set of bits with the same place value) for \emph{all} coefficients, from most to least
significant bit.  Examples originally developed for wavelet-based image compression
include embedded zerotree wavelet (\ezw) coding~\cite{ezw}, set partitioning in
hierarchical trees (\spiht)~\cite{spiht}, and variants such as set partitioned embedded
block (\speck)~\cite{speck}.  These techniques capitalize on the sparsity of wavelet
coefficients, which are often small with many leading zeros that can be encoded
efficiently together.  Using the idea of \emph{group testing}~\cite{Hong:2001}, many bits
are pooled, tested together, and then iteratively refined if the group contains at least
one one-bit.  Such techniques have been adopted in scientific data compressors like
\zfp~\cite{Li14} and \sperr~\cite{sperr}, though without direct support for progressive
reconstruction.  Also, recognizing that many zero-bits tend to occur together, long runs
of zeros may be encoded using \emph{run-length encoding} (RLE), as in the \tthresh
compressor~\cite{tthresh}.

We note that embedded coders like \ezw and \spiht are generally presented not in terms of
encoding bit planes but rather as encoding comparisons of coefficient magnitudes with a
decreasing sequence of \emph{thresholds}, $\tau_1 > \tau_2 > \tau_3 \cdots$.  A
coefficient exceeding such a threshold is deemed \emph{significant}.  Once a coefficient
has been found to be significant, its less-significant bits are coded in subsequent passes
separate from the significance tests.  When $\tau$ is an integer power of two, as is
typically the case, this approach reduces to bit plane coding.  Our progressive framework
also is based on a sequence of thresholds or error tolerances, though not necessarily
consecutive integer powers of two.

Recognizing that these two camps represent two extremes---either encode
\emph{full} precision for a subset of coefficients or encode \emph{all}
coefficients at the same precision---Hoang \etal~\cite{Hoang2019,Hoang2021}
recently proposed exploring both dimensions of progression simultaneously,
thus providing flexibility in trading precision and resolution tailored to
the task at hand.
An alternative representation with similar capabilities was more recently
proposed~\cite{amm}, with less fine-grained progression in precision.
With recent additions to \mgard~\cite{mgard}, one may similarly
refine in precision or resolution~\cite{LiGoCh21,GoWhZh22}.
Likewise, wavelet-based image compressors such as JPEG2000~\cite{jpeg2000} organize
bit streams to support refinement in both resolution and precision.
Through the remainder of this paper, however, we restrict our discussion to
progression in precision.

We note that the progressive techniques discussed here---especially those that support
progression in both precision and resolution---require additional data structures to
maintain not only the current approximation in the progression but also sufficient
information to allow refining that approximation---i.e., the current ``cut''---represented
for example as hash maps of grid points~\cite{amm} or partial wavelet
sub-bands~\cite{Hoang2021}.  This auxiliary information may also include partial state,
such as which coefficients are significant~\cite{ezw,spiht,Hoang2021}, to allow the
decompressor to know how to interpret the next sequence of bits once they arrive.
Moreover, custom file formats are needed that support indexing for random access to small
chunks of data~\cite{Hoang2021}, and possibly precomputed \emph{error
matrices}~\cite{LiGoCh21} that capture the errors resulting from any given combination of
resolution and precision.  We avoid this additional complexity by decoupling the
\emph{components} that serve as units of refinement, which are (de)compressed
independently using an error-bounded compressor.

We conclude this section by referencing the inspiration behind our progressive framework:
\emph{multi-component representations}~\cite{Dekker:1971,Shewchuk:1997,Hida:2001}.
Originally developed as a means to extend floating-point precision beyond
hardware-supported types, a multi-component representation uses a vector of low-precision
numbers to approximate a higher-precision number, wherein each vector component represents
the remaining rounding error in the approximation associated with the sum of previous
components (see the subsequent section for details).  The same principle also underpins
the family of \emph{iterative refinement} methods for linear
solvers~\cite{Wilkinson:1963}, where one solves not just for a single vector but also a
hierarchy of error vectors to boost accuracy, as well as the recently proposed \hire
algorithm~\cite{Barbarioli23}, which targets progression in resolution.  To our knowledge,
our approach is the first to generalize these ideas from multi-component floating-point
representation for high-precision arithmetic to the field of lossy data compression.

%% file: preliminaries.tex
\section{Preliminaries}
\label{sec:preliminaries}

Before describing our approach, we review the idea behind multi-component
representations that underpin it.  The key idea is to represent a high-precision
number using a vector of lower-precision components, with each additional
component capturing the rounding error achieved so far in the sum---computed
at full precision---of prior components.  In this manner, a double-precision
number with 53 mantissa bits can be well-approximated as two single-precision
numbers with 24-bit mantissas by effectively concatenating their mantissas.
However, note that single precision has smaller dynamic range, limiting
the domain of values that can be represented this way.

Let $\fl[\cdot]$ represent the operator that rounds its argument to the nearest
low-precision floating-point number.  Given a high-precision number, \actual,
we define the first component as $\comp{1} = \fl[\actual]$.  This results in a
rounding error $\error = \actual - \comp{1}$, which is captured in a second component:
$\comp{2} = \fl[e] = \fl[x - x_1]$ such that $x \approx x_1 + x_2$, where the
addition is computed in full precision (the precision used for $x$).
Note that this holds only approximately since there is also a secondary
rounding error in the computation of $x_2$.  If equality is not achieved,
a third component may be introduced: $x_3 = \fl[x - (x_1 + x_2)]$, and so
on, with $x \approx \sum_i x_i$ and $\comp{k} = \fl[x - \sum_{i=0}^{k-1} \comp{i}]$.
By convention, $\comp{0} = 0$.  Note that because floating-point addition
is not guaranteed associative, the order in which components are added matters.
For consistency, terms are added eagerly from left to right, with each additional
component appended to the previous sum:
$\csum{k+1} = \csum{k} + \comp{k+1} = \bigl(\sum_{i=0}^{k} \comp{i}\bigr) + \comp{k+1}$.
Since $\comp{k+1} = \fl[\actual - \csum{k}]$, $|\comp{k+1}| \approx |\actual - \csum{k}|$
represents an estimate (accurate to machine epsilon) of the error in using
\csum{k} to approximate \actual.

%% file: method.tex
\section{Multi-Component Representation}
\label{sec:multicomp}

Equipped with a multi-component construction algorithm for scalars based on
conventional floating point, we now generalize this approach to scalar fields,
\vactual, stored in other number representations.  In place of \fl[\cdot], we rely
on lossy compression, $\func{C}(\cdot)$, and decompression, $\func{D}(\cdot)$,
functions, with the composition $\func{D} \circ \func{C}$ taking on the role
previously held by \fl[\cdot] to approximate the input using reduced precision.
Any lossy compression algorithm takes some parameter that governs the amount of
loss.  This could, for instance, be a target bit rate that directly controls the
amount of compressed storage~\cite{Li14}, a target error under some chosen
norm, like $L_2$~\cite{tthresh}, or an error tolerance, \tol, that bounds the
pointwise maximum absolute difference between the uncompressed and lossy-compressed
fields, as supported by most contemporary compressors~\cite{mgard,sperr,LiZhSh23,Li14}.
We will present our framework in terms of a decreasing sequence of such absolute error
tolerances, $\tol_1 > \tol_2 > \cdots > \tol_n$, while noting that other compression
parameters can easily be substituted.

\subsection{Progressive Compression}

\begin{algorithm}[!t]
\caption{Multi-component data compression}
\begin{algorithmic}[1]
\REQUIRE $\vactual;\, n;\, \func{C}(\vcomp{}, \tol);\, \func{D}(\vzcomp{});\, (\tol_1, \tol_2, \ldots, \tol_n) : \tol_{i} > \tol_{i+1}$
\ENSURE $(\vzcomp{1}, \vzcomp{2}, \ldots, \vzcomp{n})$
\STATE $\vapprox \leftarrow \vec{0}$
\FOR{$i \gets 1, \ldots, n$}
  \STATE $\verror \leftarrow \vactual - \vapprox$
  \STATE $\vzcomp{i} \leftarrow \func{C}(\verror, \tol_{i})$
  \STATE $\vcomp{i} \leftarrow \func{D}(\vzcomp{i})$
  \STATE $\vapprox \leftarrow \vapprox + \vcomp{i}$
\ENDFOR
\end{algorithmic}
\label{alg:multicomp1}
\end{algorithm}

\Cref{alg:multicomp1} details our multi-component construction method.  The method takes
as input the desired number of components, $n$, the original (uncompressed) data as a
vector of double-precision floating-point values, \vactual, and the aforementioned
sequence of decreasing error tolerances, $(\tol_1, \tol_2, \ldots, \tol_n)$, selected by
the user.  For the sake of simplicity, we omit other input parameters that are needed by
the compressor such as the grid dimensions of the uncompressed field.  On Line~1, the
current approximation, \vapprox, to \vactual is initialized to a field of all zeros.
Next, we loop over the components, \vcomp{i}, to be constructed.  In each iteration, we
compute the current error $\verror = \vactual - \vapprox$, which decreases with each
iteration.  \verror is then compressed to within a tolerance, $\tol_i$, forming the
$i^\text{th}$ compressed component, \vzcomp{i}.  We then immediately decompress
\vzcomp{i}, yielding \vcomp{i}.  Given the imposed error tolerance, we are thus assured
that $\linf{\verror - \vcomp{i}} \leq \tol_i$.  Finally, we update \vapprox by adding in
the just computed component, \vcomp{i}, resulting in a more accurate approximation.  The
output of the algorithm is the sequence of compressed components, $(\vzcomp{1},
\vzcomp{2}, \ldots, \vzcomp{n})$.

\subsection{Progressive Reconstruction}

\begin{algorithm}[!t]
\caption{Multi-component data reconstruction}
\begin{algorithmic}[1]
\REQUIRE $m \leq n;\, \func{D}(\vzcomp{});\, (\vzcomp{1}, \vzcomp{2}, \ldots, \vzcomp{n})$
\ENSURE $\vapprox$
\STATE $\vapprox \leftarrow \vec{0}$
\FOR{$i \gets 1, \ldots, m$}
  \STATE $\vcomp{i} \leftarrow \func{D}(\vzcomp{i})$
  \STATE $\vapprox \leftarrow \vapprox + \vcomp{i}$
\ENDFOR
\end{algorithmic}
\label{alg:multicomp2}
\end{algorithm}

A progressive reconstruction based on a subset of $m \leq n$ components is obtained by
executing \cref{alg:multicomp2}.  This algorithm is essentially the same as
\cref{alg:multicomp1} but with Lines~3--4, which compute and compress the error, excluded,
and with the loop on Line~2 shortened to $m$ iterations.  We note that we can trivially
continue refining the reconstruction, \vapprox, by incorporating additional components as
they arrive by simply executing additional loop iterations.

\subsection{Error Bounds}
\label{sec:error-bounds}

The sequence of error tolerances utilized by our construction algorithm serves not only to
induce a sequence of approximations but also as (approximate) error bounds in the
resulting reconstructions $\{\vapprox_i\}$.  To see why this is, let $\verror[i] =
\vactual - \vapprox_{i-1}$.  As the single-component compressor $\func{C}$ ensures
$\linf{\verror[i] - \vcomp{i}} \leq \tol_i$, we have
\begin{align}
\begin{split}
     \linf{\vactual - \vapprox_i} 
  &= \linf{\vactual - (\vapprox_{i-1} + \vcomp{i})} \\
  &= \linf{(\vactual - \vapprox_{i-1}) - \vcomp{i}} 
   = \linf{\verror[i] - \vcomp{i}} 
   \leq \tol_i.
\end{split}
\end{align}
However, we here rely on the use of the associative rule $\vactual - (\vapprox_{i-1} +
\vcomp{i}) = (\vactual - \vapprox_{i-1}) - \vcomp{i}$, which does not necessarily hold as
computations are being performed in floating-point arithmetic, i.e., this bound may be
violated by a small constant proportional to machine epsilon.  This is in practice of
little consequence as \vactual already is contaminated by similar rounding errors, which
in turn are often dwarfed by larger sensor, truncation, iteration, and model errors.

In case the underlying compressor is not driven by an absolute error tolerance but by some
other compression parameter, we advocate \emph{computing} the corresponding errors, i.e.,
set $\tol_i \gets \linf{\vactual - \vapprox_i}$ and maintain $\{\tol_i\}$ as additional
metadata for later use in specifying the required accuracy.  One may even choose to do so
with error-bounded compressors that provide only a loose error bound, like \mgard and
\zfp, to further tighten the error bounds.

\subsection{Compressibility of Components}

We note that in each iteration of our algorithm, we are compressing the remaining
\emph{error}, $\error[i]$, in the approximation.  As $i$ increases, these errors tend to
become increasingly random as any spatial correlation (or ``smoothness'') in the data is
lost.  Such randomness is usually a bane for numerical compressors, and we would expect
compression to become increasingly less effective.  Worse yet, due to the pigeonhole
principle, some inputs to a compressor must lead to \emph{expansion}, so it is legitimate
to question the efficacy of our approach.

We make two points in defense of our framework: (1)~Whether one compresses a data set in
batches using multiple components or in a single sweep, the same high-entropy
less-significant bits that are still needed to attain a certain level of accuracy must be
faithfully preserved.  In a conventional single-component compressor, such trailing bits
are either encoded verbatim explicitly, as in~\cite{Li14,spiht,speck,sperr}, or
implicitly, e.g., using Huffman codes~\cite{LiZhSh23}.  In a multi-component compressor,
groups of consecutive bit planes are effectively isolated\footnote{
Line~3 in \cref{alg:multicomp1} effectively peels off leading bits already encoded in
prior components, exposing less significant bits.}  and compressed independently, however
this in and of itself does not significantly impact the compressibility of said bits.
Rather, the main penalty in using a multi-component representation lies in the
per-component overhead one may expect, e.g., \zfp encodes one exponent per block and
component; \sz embeds a Huffman table with each component; and \sperr encodes a
significance map with each component.  However, as we shall see, compressing the data in
batches (components) can also \emph{improve} both compression and error.  (2)~As precision
increases, the magnitude of components decreases, until eventually many values are exactly
represented with zero error.  As components (i.e., errors) become increasingly sparse,
they also become more compressible.  We will evaluate the effects that randomness plays in
the following section.

%% file: results.tex
\section{Numerical Results}
\label{sec:results}

In this section, we evaluate our multi-component approach using real-world data fields,
primarily from SDRBench~\cite{ZhDiLi20}; see \cref{tab:data}.  We use four
state-of-the-art error-bounded lossy compressors (see \cref{tab:compressors}) within the
context of our framework, representing a range of different compression techniques:
\begin{itemize}
   \item \zfp is a block-transform-based compressor (similar to JPEG) that uses custom bit
     plane coding.

   \vspace{-1ex}
   \item \sz encodes corrections to predictions that violate the error tolerance using
     Huffman coding.

   \vspace{-1ex}
   \item \sperr is based on CDF~9/7 wavelets. Like \sz, it encodes corrections to
     outliers, however using \speck.

   \vspace{-1ex}
   \item \mgard uses a custom multilevel basis akin to CDF~5/3 wavelets, coupled with
     quantization and lossless compression.
\end{itemize}
We conducted the numerical experiments on a workstation equipped with 64~GB of RAM and an
Intel Xeon E5-1650v4 CPU. All source codes were compiled with GCC~12.2.1 and the
optimization flags \texttt{-O3 -march=native}. We note that many of the data compressors
tested in this work have support for multi-threading or GPU acceleration. However, we
decided to test only the single-core CPU execution mode to simplify our analysis.

\begin{table}[!tb]
\caption{Data sets used to benchmark our approach.}
\centering
\scriptsize
\begin{tabular}{llr@{\ $\times$\ }r@{\ $\times$\ }rc}
\toprule
Name                        & Application type &
\multicolumn{3}{c}{Dimensions} & Precision \\
\midrule
Miranda   & Hydrodynamics     &  384 &  384 & 256 & 64 \\
S3D       & Combustion        &  500 &  500 & 500 & 64 \\
S3D JICF~\cite{jicf}
          & Combustion        &  400 &  250 & 200 & 64 \\
\midrule
Nyx       & Cosmology         &  512 &  512 & 512 & 32 \\
QMCPACK   & Quantum chemistry &   69 &   69 & 115 & 32 \\
SCALE     & Weather           & 1200 & 1200 &  98 & 32 \\
\bottomrule
\end{tabular}
\label{tab:data}
\end{table}

\begin{table}[!tb]
\caption{Compressors used in our multi-component framework (top) and
compared against (bottom).}
\tabcolsep 3pt
\centering
\scriptsize
\begin{tabular}{lcl}
\toprule
Compressor  & Version/Hash & GitHub repository \\
\midrule
\zfp~\cite{Li14}         & $1.0.0$   & \scriptsize{\github{LLNL/zfp}} \\
\szthree~\cite{LiZhSh23} & $3.1.7$   & \scriptsize{\github{szcompressor/SZ3}} \\
\sperr~\cite{sperr}      & $0.6.2$   & \scriptsize{\github{NCAR/SPERR}} \\
\mgard~\cite{mgard}      & $28d738c$ & \scriptsize{\github{lxAltria/MGARDx}} \\
\midrule
\idxtwo~\cite{Hoang2021} & $4f072cb$ & \scriptsize{\github{sci-visus/idx2}} \\
\pmgard~\cite{LiGoCh21}  & $e581794$ & \scriptsize{\github{lxAltria/Multiprecision-data-refactoring}} \\
\fpzip~\cite{fpzip}      & $1.3.0$   & \scriptsize{\github{LLNL/fpzip}} \\
\bottomrule
\end{tabular}
\label{tab:compressors}
\end{table}

\subsection{Error Analysis}
\label{subsec:results_error}

We begin by evaluating the rate-distortion associated with the four compressors discussed
above when (1)~used in isolation, which we will refer to as ``single-component
compressors,'' and when (2)~incorporated into our multi-component framework.  Toward this
end, we used a sequence of absolute error tolerances, $\tol_i = 2^{-\gran \times i}
\tol_0$, during compression, with $\tol_0$ representing the range of field values and
$\gran \in \{4, 6, 8\}$ being the \emph{granularity} of progression, in essence governing
the precision associated with each component.  We used $\gran = 2$ with the
single-component compressors and reran them from scratch to obtain a data point for each
corresponding error tolerance.

\begin{figure}[tp]
\includegraphics[width=.5\linewidth]{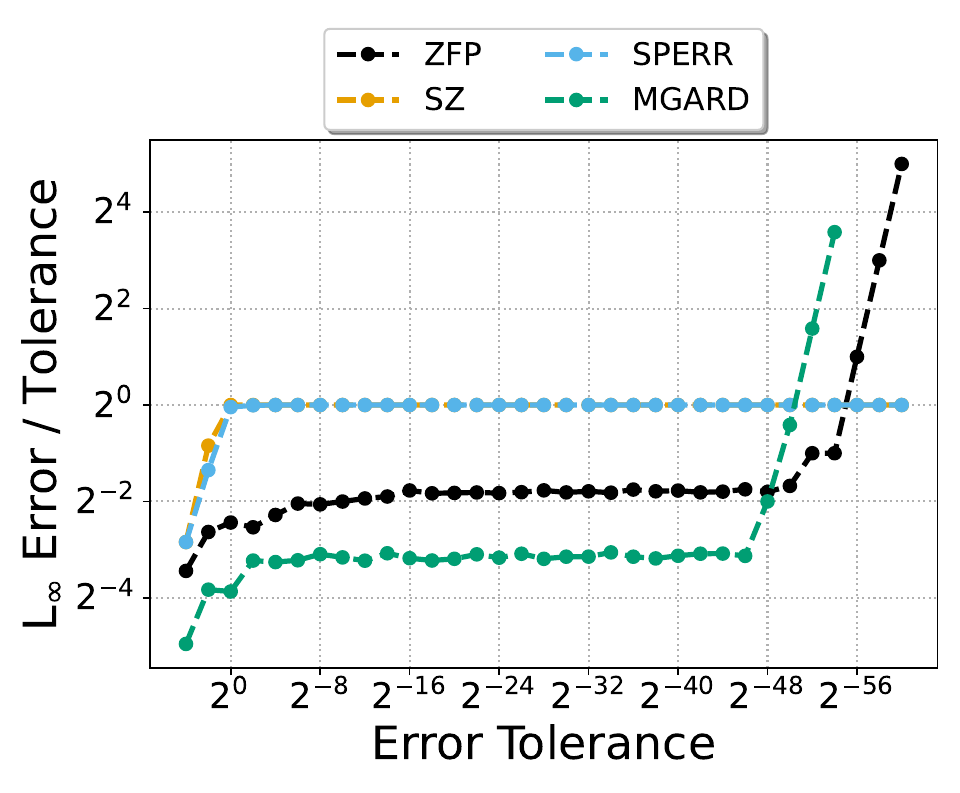}\hfill%
\includegraphics[width=.5\linewidth]{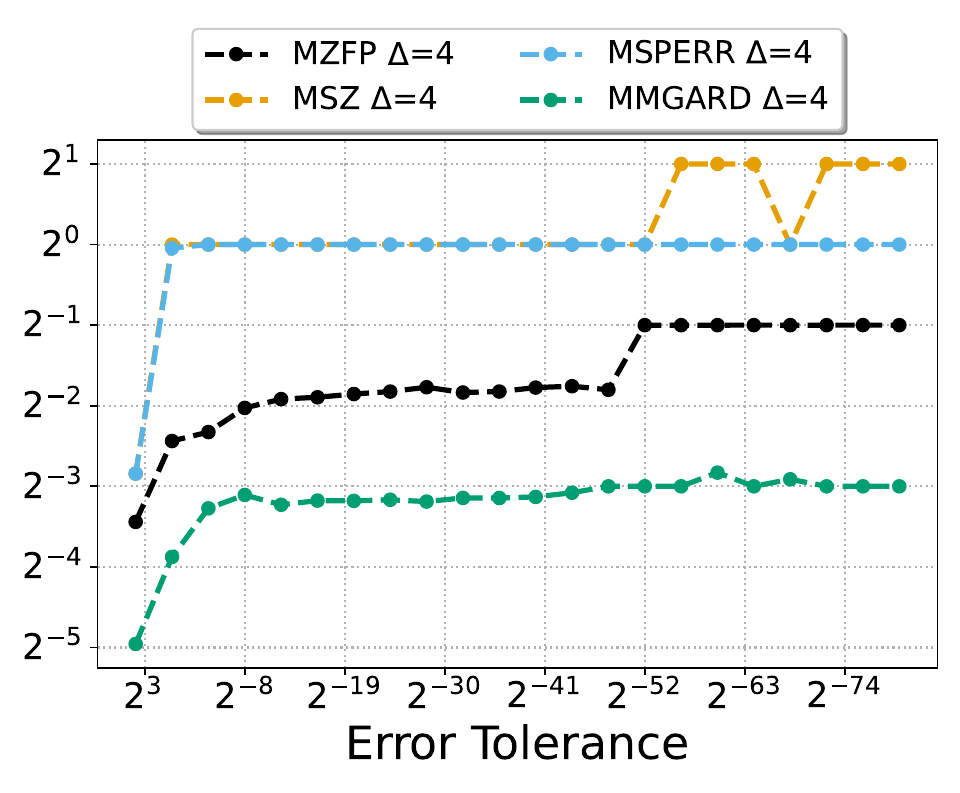}%
 \caption{Ratio of observed maximum absolute error to error tolerance,
     \tol, as a function of \tol, for the single-component
     compressors (left) and for $m$-component ($1 \leq m \leq n$) reconstructions based on
     \zfp, \sz, \sperr, and \mgard (right). Note that the observed errors generally
     respect the tolerance, either exactly, by coinciding with the tolerance
     (ratio equals $2^0 = 1$), or conservatively, by falling below the tolerance.
     See the text for a discussion of the \mgard and \zfp plots.}
 \label{fig:results1}
\end{figure}


To verify that our framework meets error bounds, we plot in \cref{fig:results1}
for the Miranda pressure field the
ratio between the measured maximum absolute error and the tolerance, \tol,
requested during progressive reconstruction, as a function of \tol. We note that the
multi-component compressors, prefixed with an `M' and suffixed with \gran in these plots,
generally respect the tolerance as this ratio tends to be at or below one.
The only exception is \msz, which at very small (relative) tolerances below machine epsilon
($2^{-52}$) occasionally gives a 1-ulp (unit in the last place) rounding error due to
the non-associativity of floating-point arithmetic, as discussed in \cref{sec:error-bounds}.
The single-component compressors (no prefix) also respect the tolerance, except for \zfp
(a known limitation~\cite[\S{4.2}]{Diffenderfer19}) and \mgard (as reported
elsewhere~\cite[\S{VI.C}]{sperr}) when \tol is near machine epsilon.
Note that \mzfp and \mmgard, our multi-component variants, overcome this limitation.

\begin{figure*}[t!]
 \includegraphics[width=.25\textwidth]{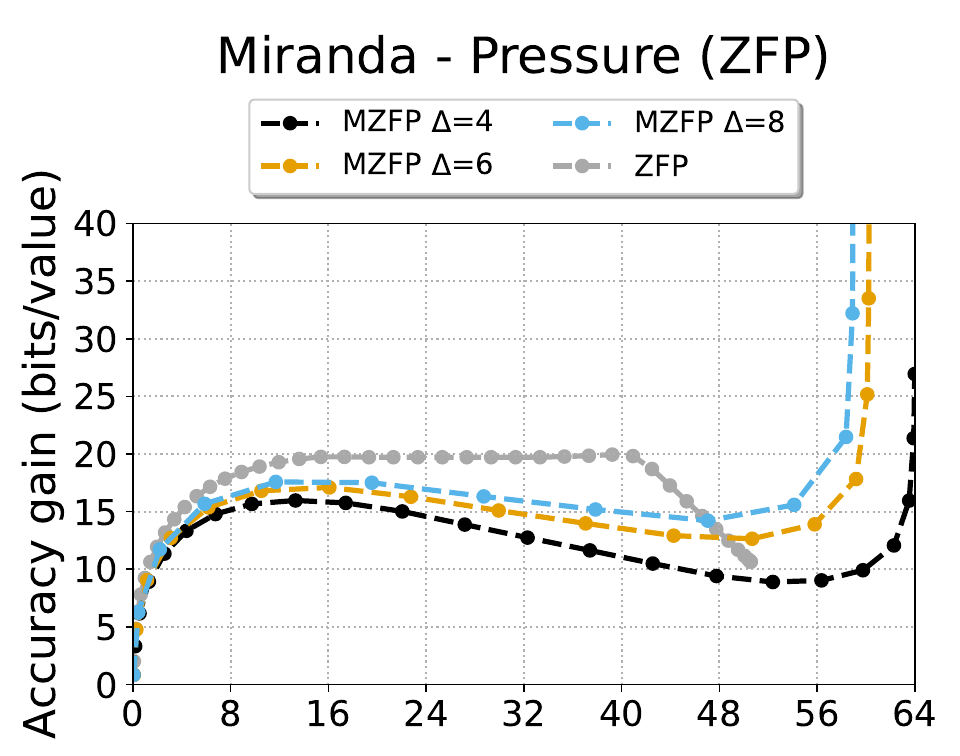}\hfill%
 \includegraphics[width=.25\textwidth]{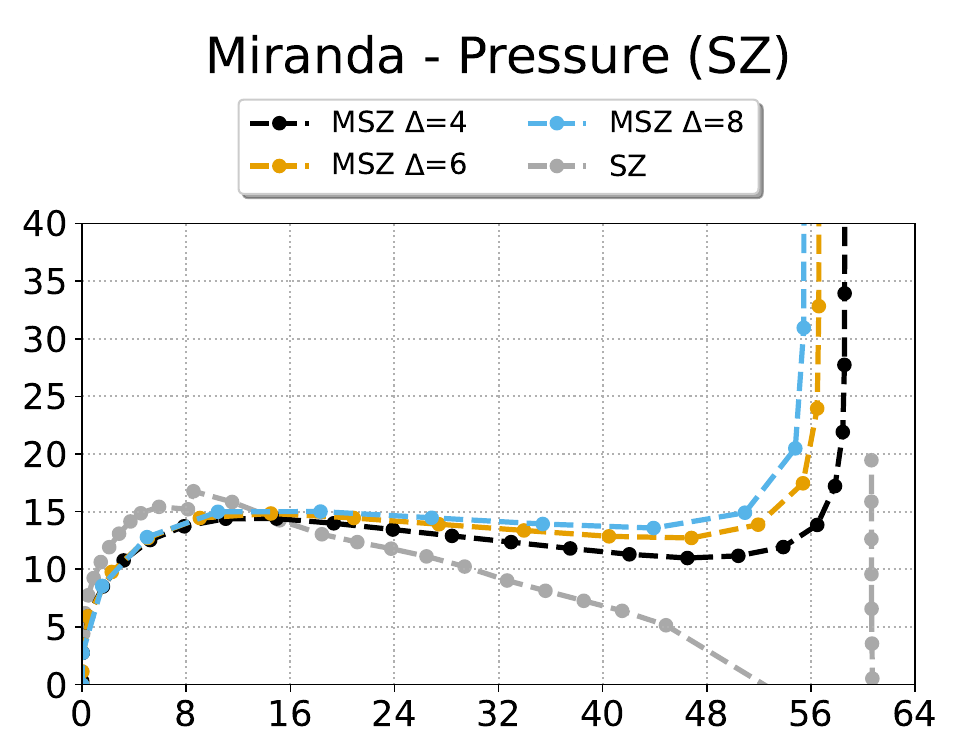}\hfill%
 \includegraphics[width=.25\textwidth]{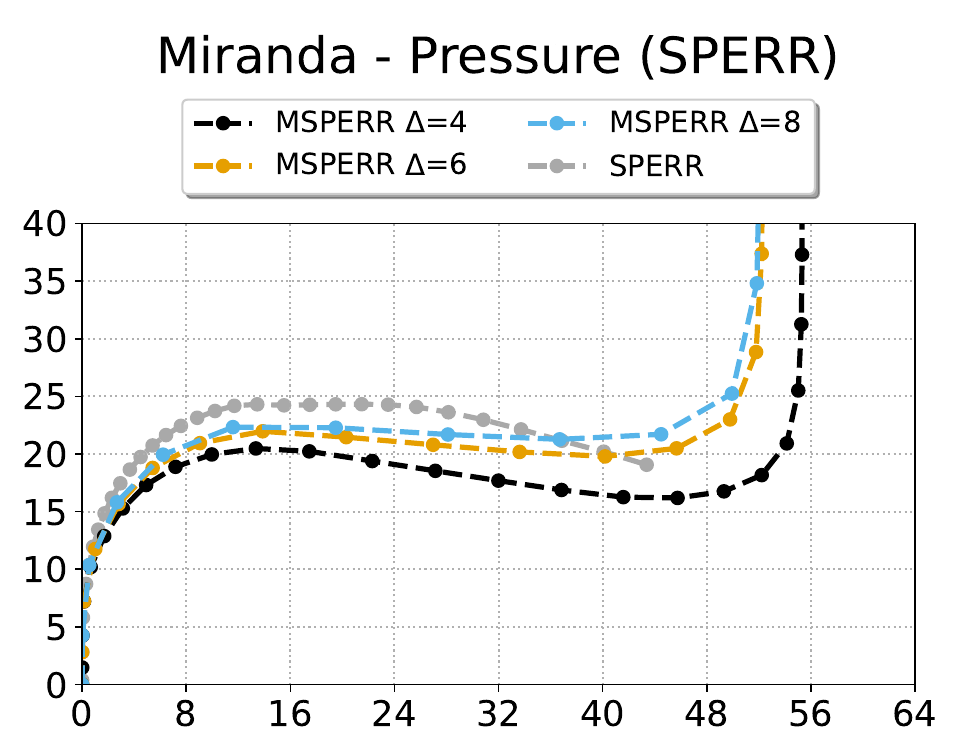}\hfill%
 \includegraphics[width=.25\textwidth]{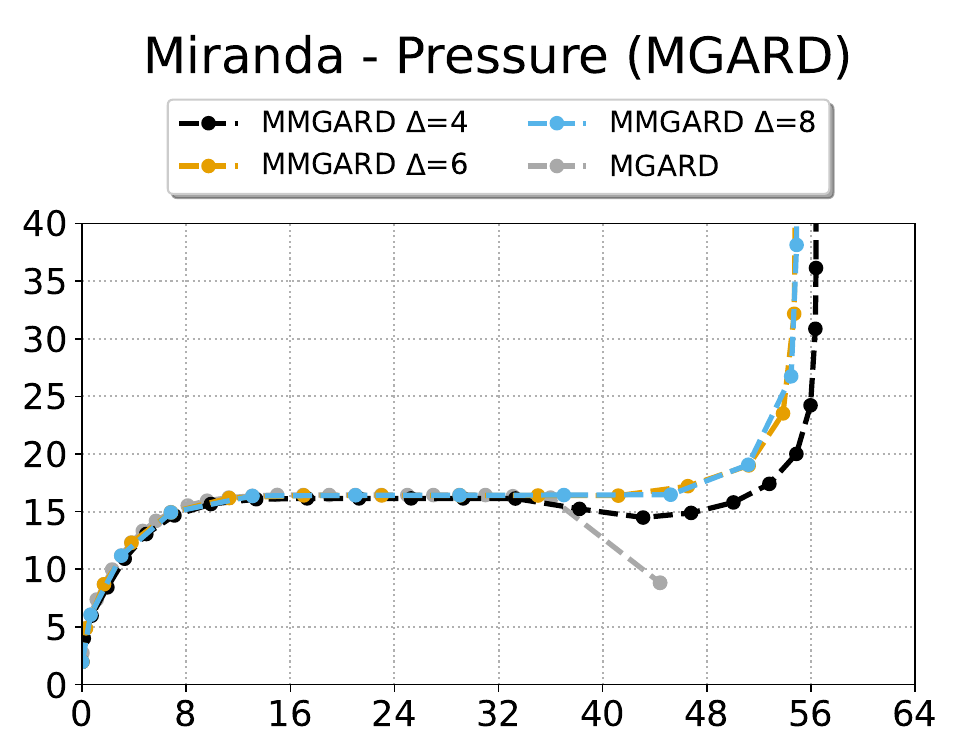}\\[-1pt]%
 \includegraphics[width=.25\textwidth]{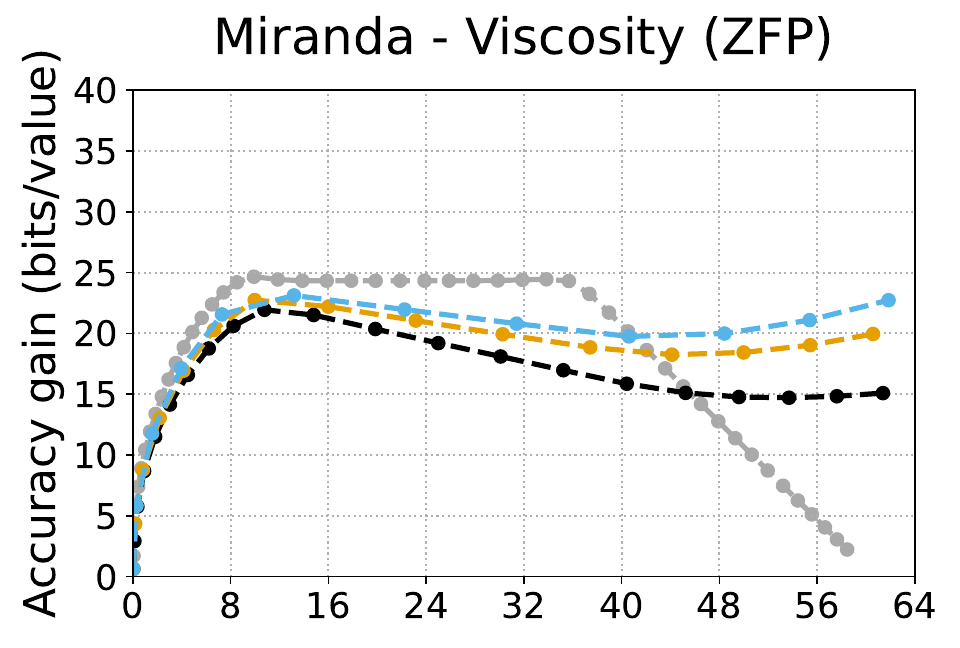}\hfill%
 \includegraphics[width=.25\textwidth]{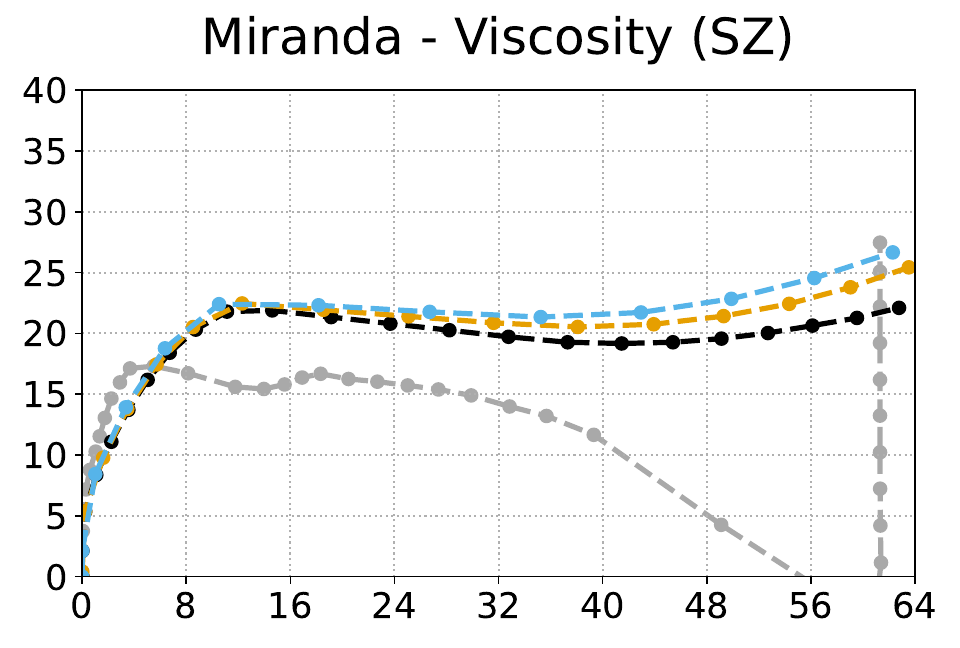}\hfill%
 \includegraphics[width=.25\textwidth]{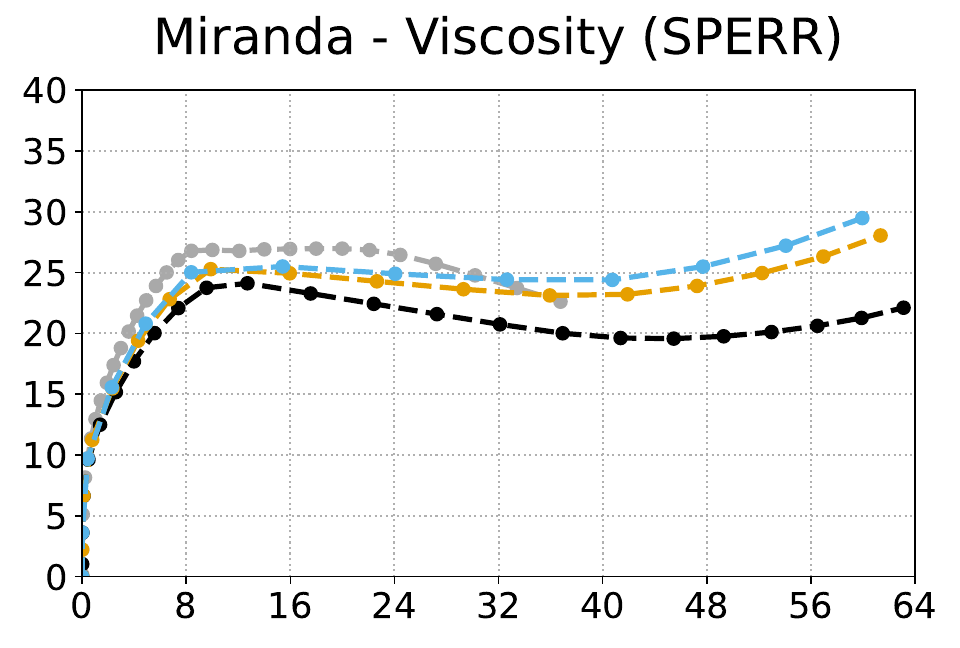}\hfill%
 \includegraphics[width=.25\textwidth]{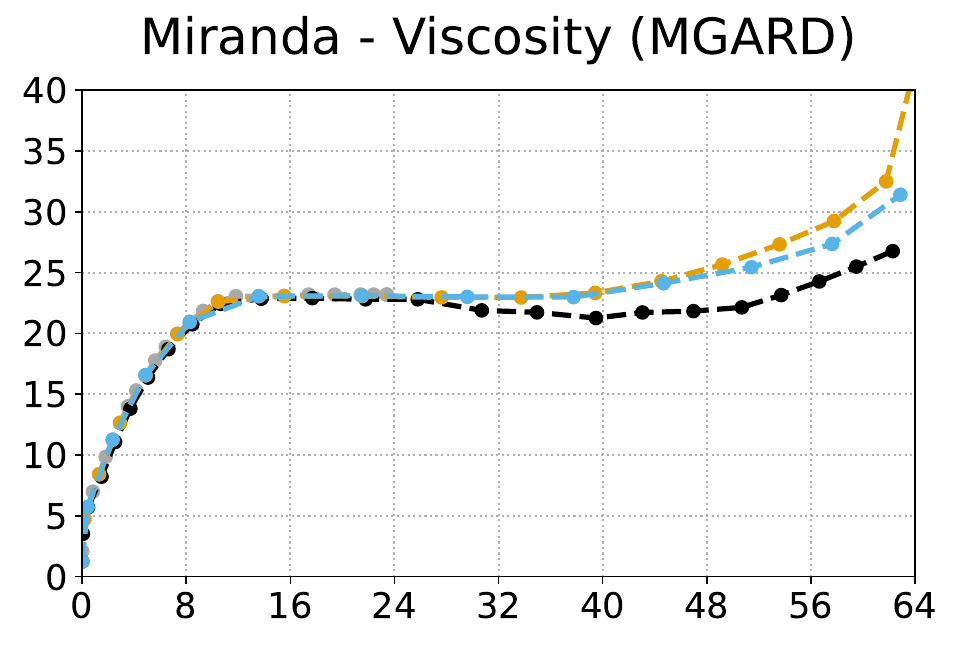}\\[-1pt]%
 \includegraphics[width=.25\textwidth]{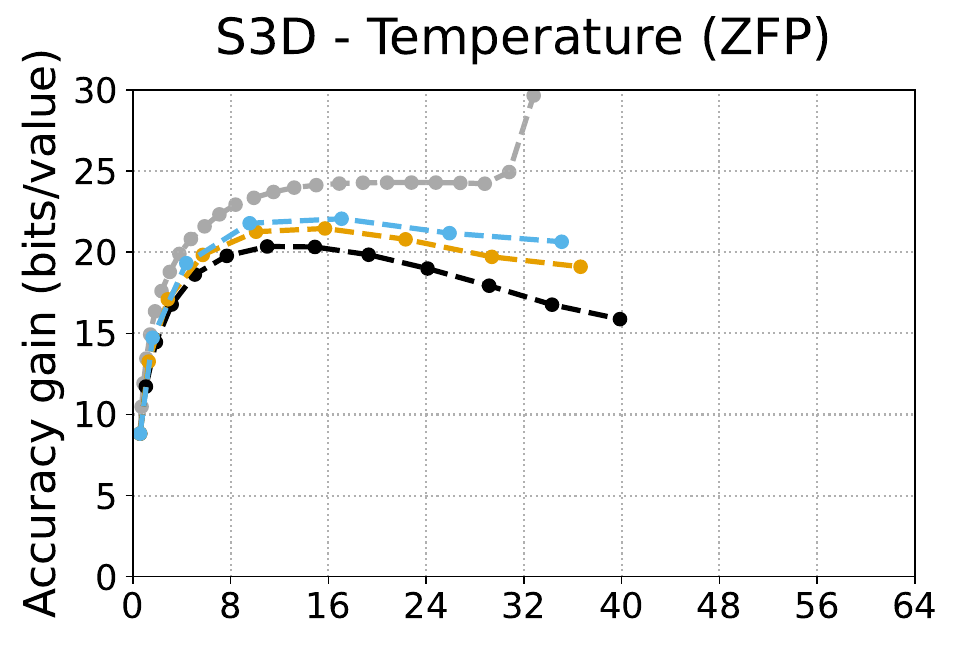}\hfill%
 \includegraphics[width=.25\textwidth]{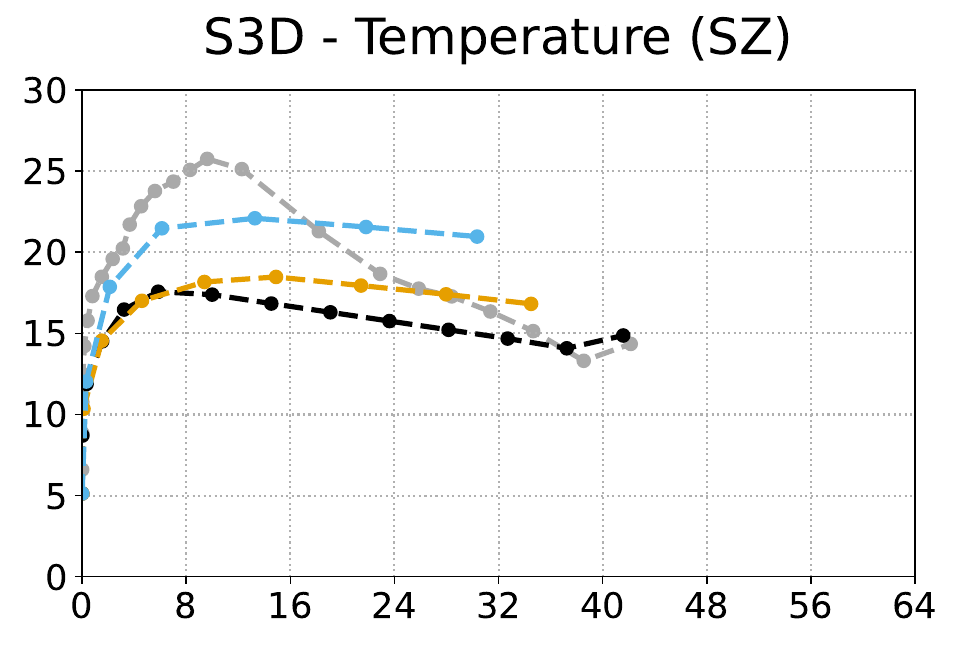}\hfill%
 \includegraphics[width=.25\textwidth]{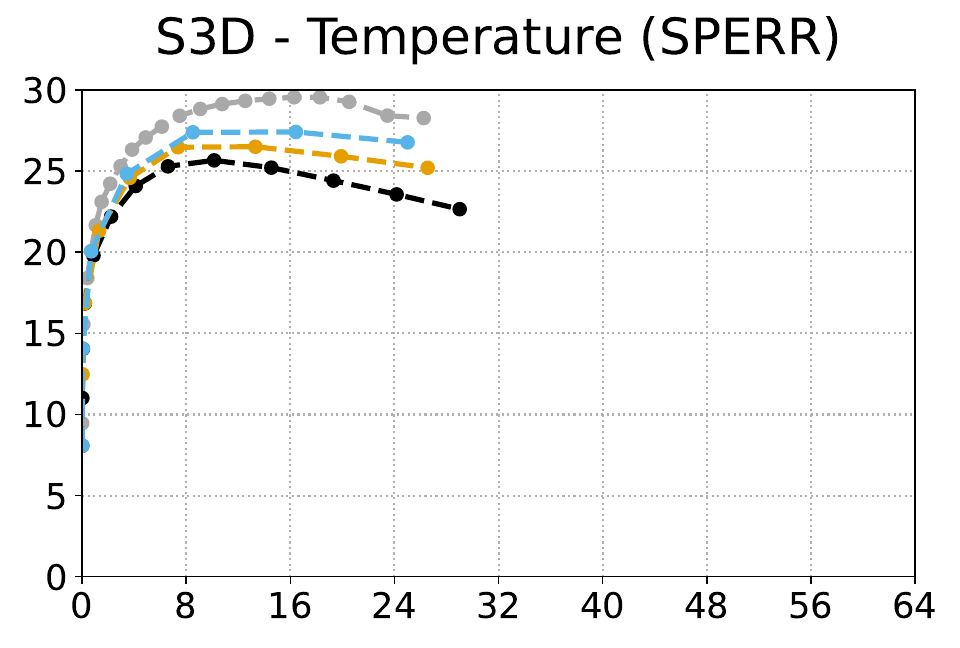}\hfill%
 \includegraphics[width=.25\textwidth]{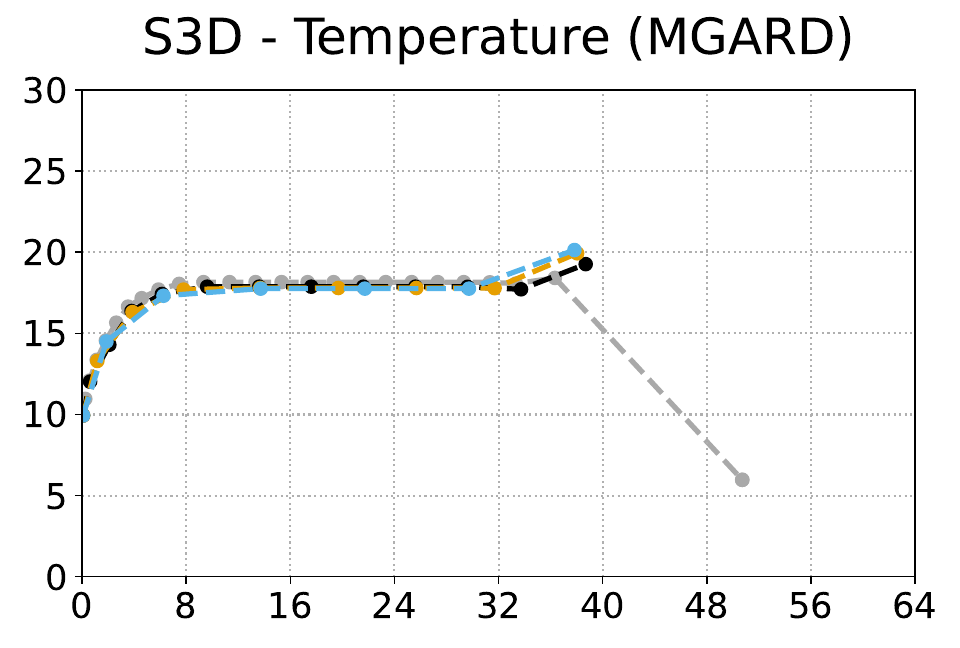}\\[-1pt]%
 \includegraphics[width=.25\textwidth]{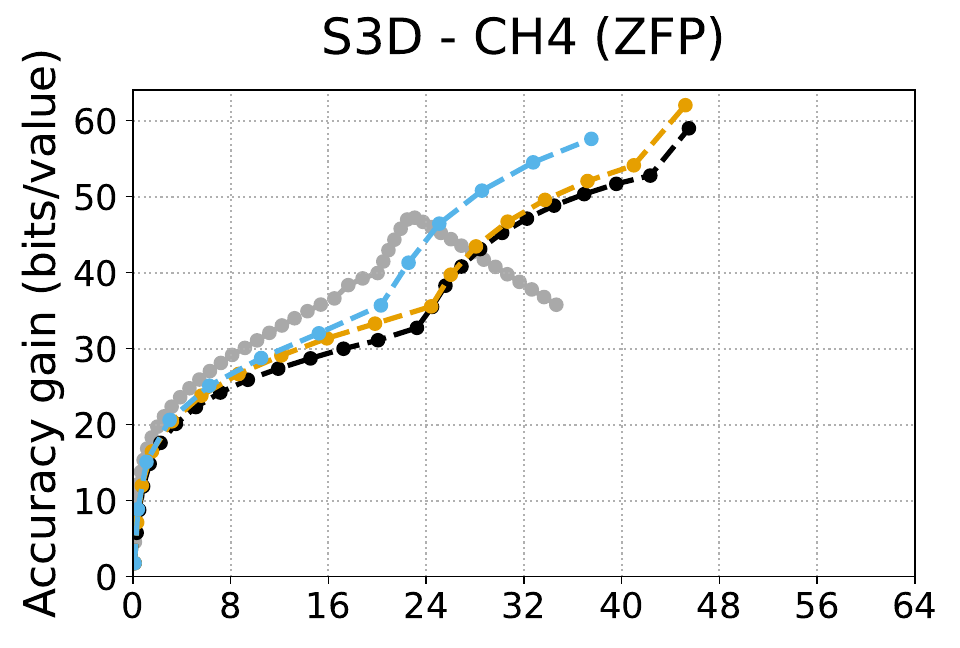}\hfill%
 \includegraphics[width=.25\textwidth]{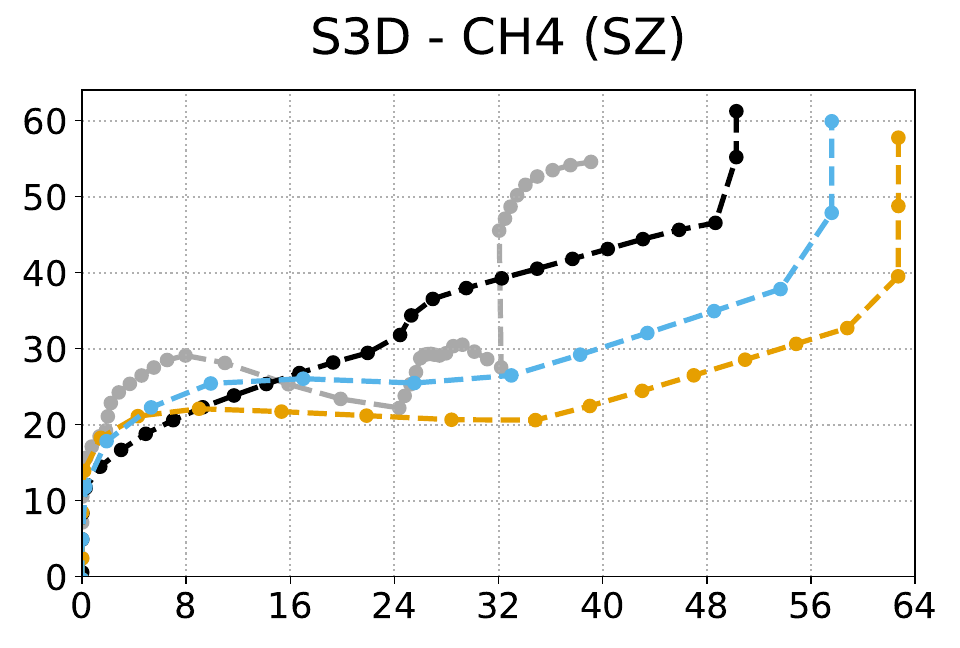}\hfill%
 \includegraphics[width=.25\textwidth]{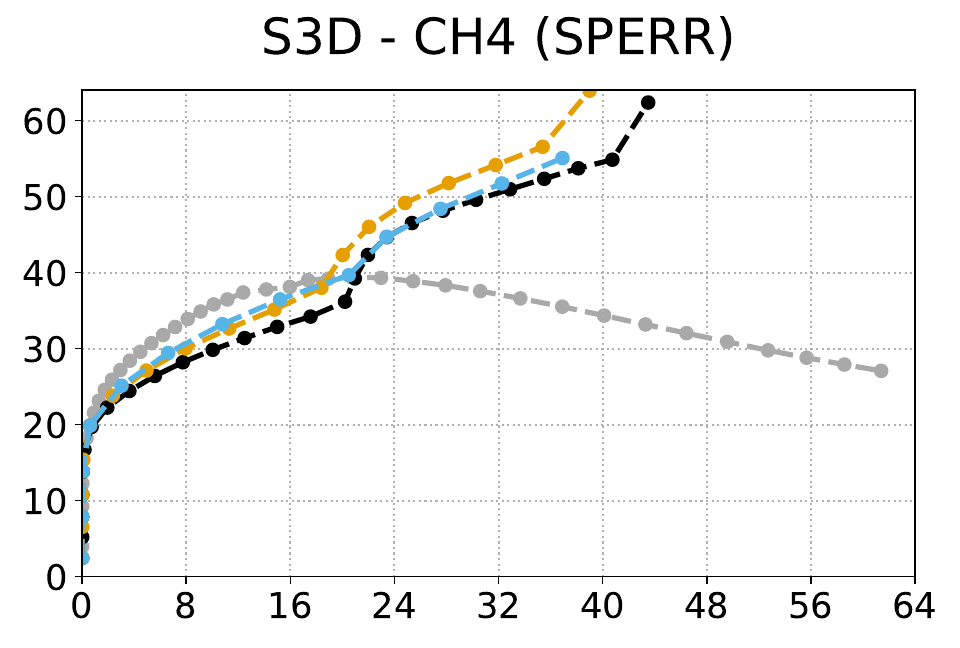}\hfill%
 \includegraphics[width=.25\textwidth]{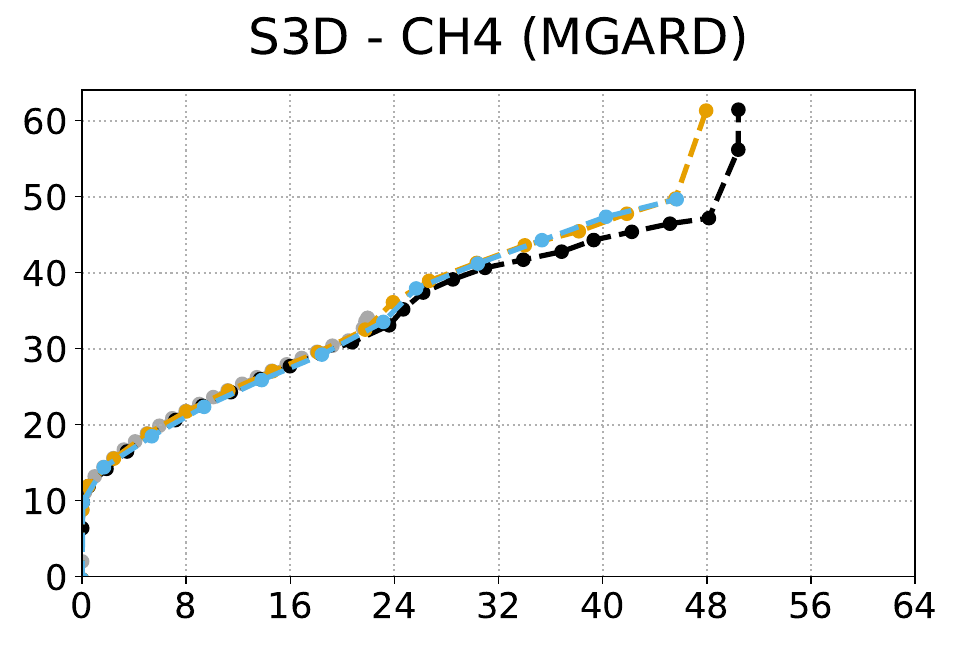}\\[-1pt]%
 \includegraphics[width=.25\textwidth]{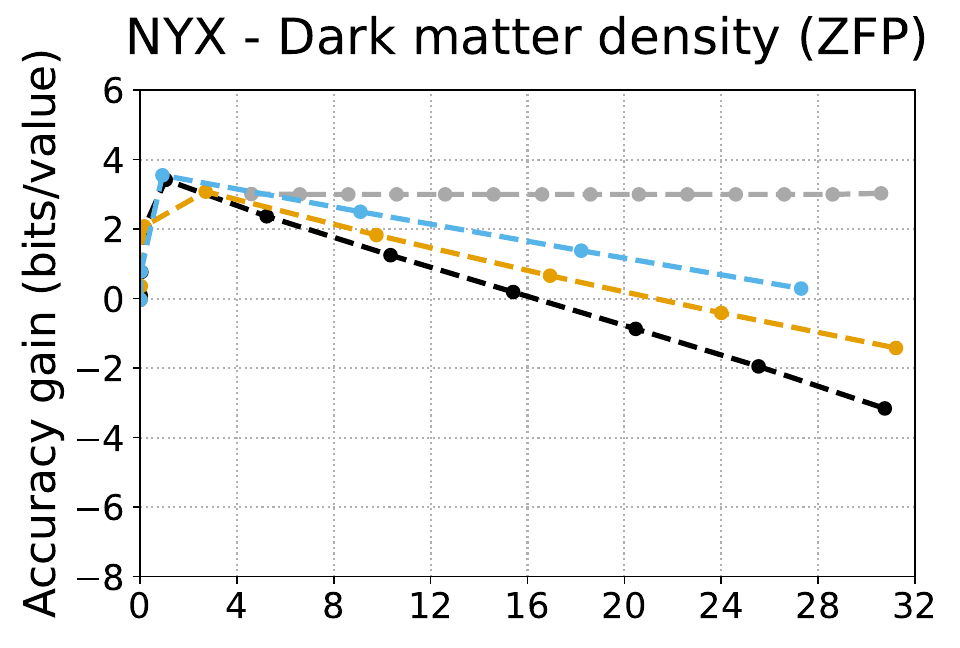}\hfill%
 \includegraphics[width=.25\textwidth]{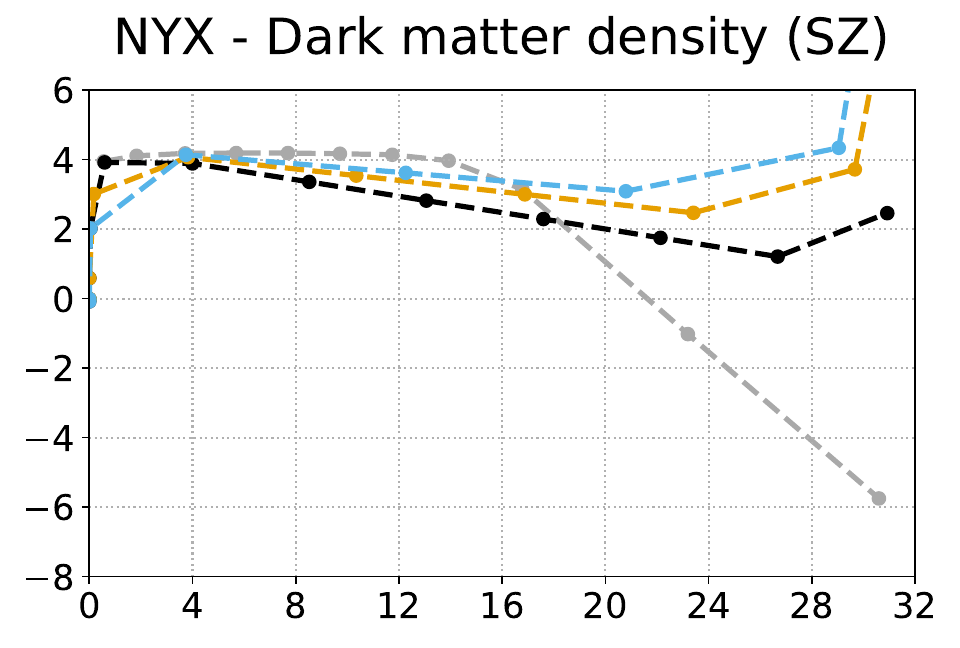}\hfill%
 \includegraphics[width=.25\textwidth]{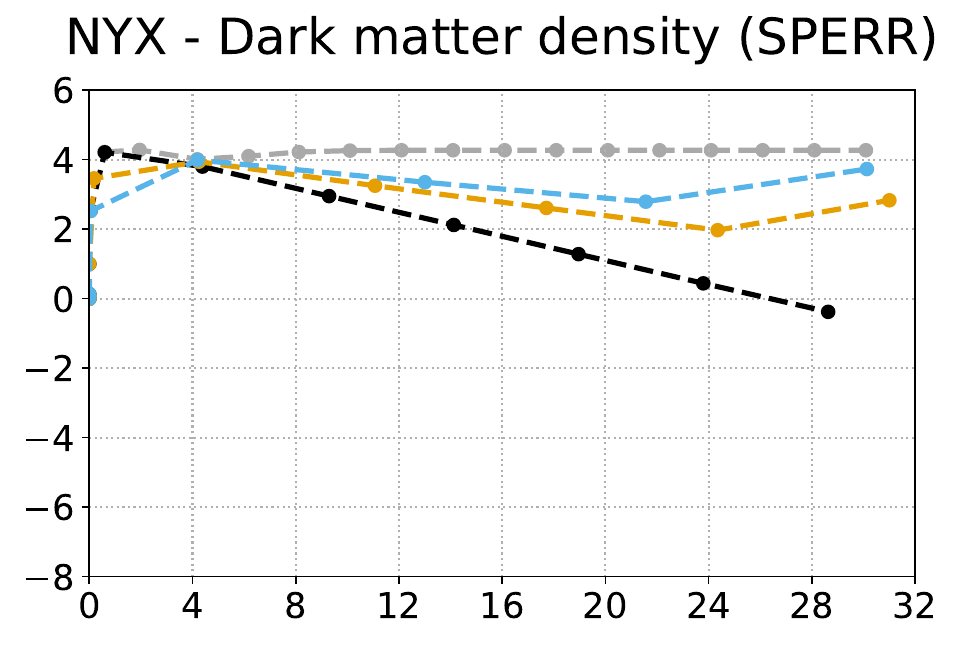}\hfill%
 \includegraphics[width=.25\textwidth]{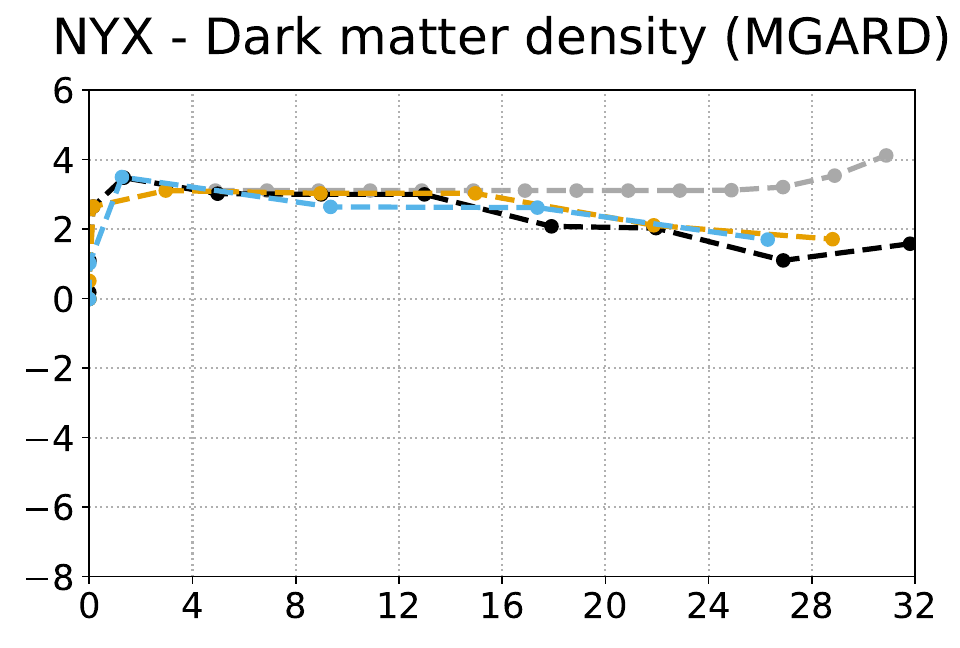}\\[-1pt]%
 \includegraphics[width=.25\textwidth]{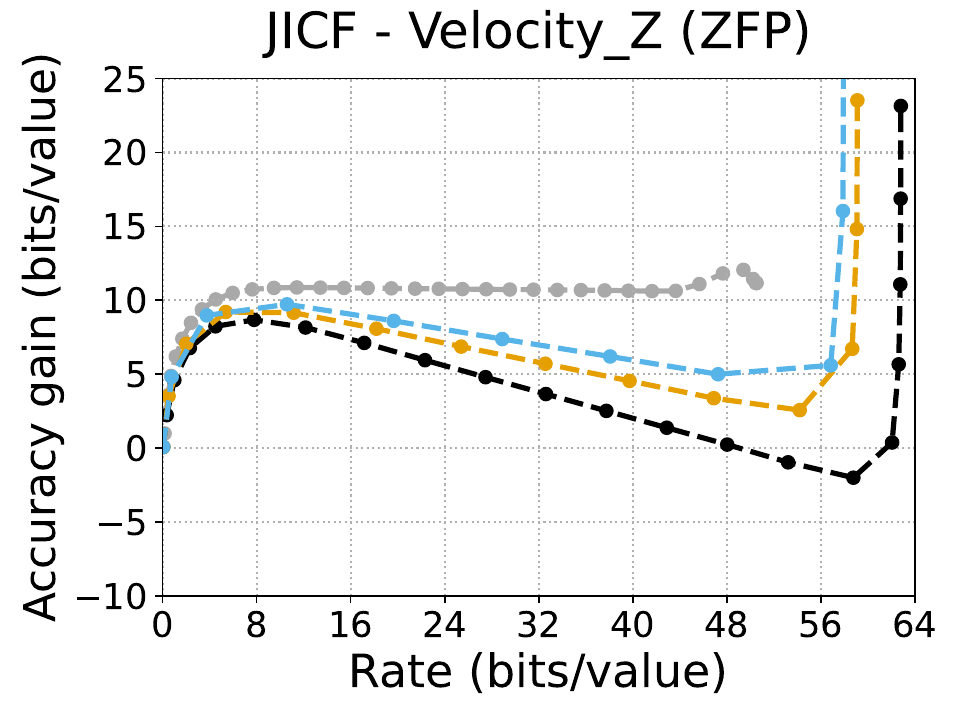}\hfill%
 \includegraphics[width=.25\textwidth]{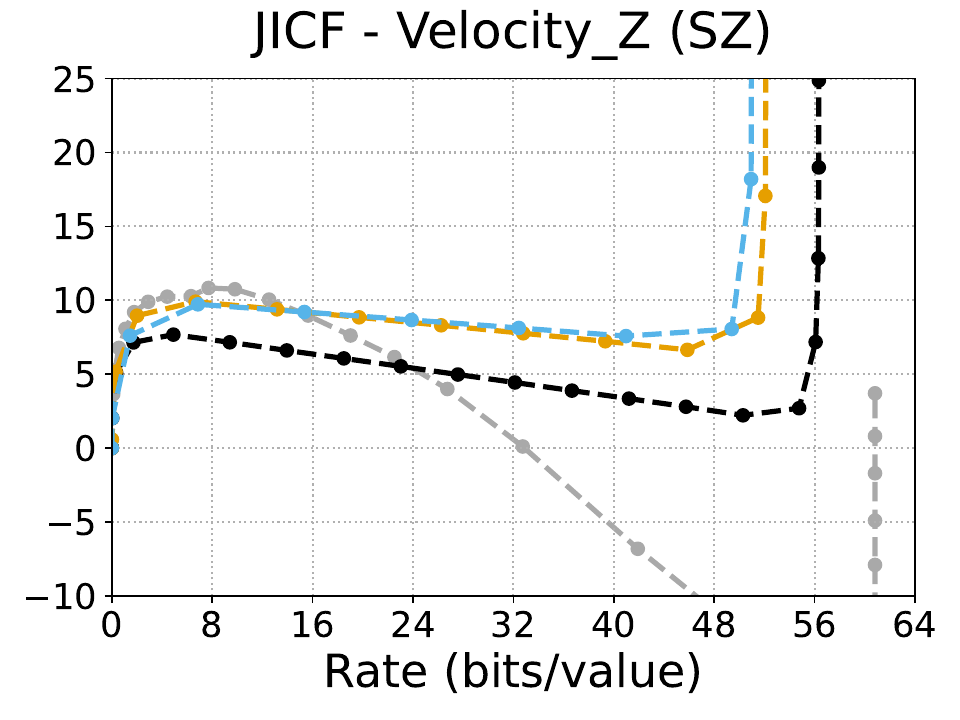}\hfill%
 \includegraphics[width=.25\textwidth]{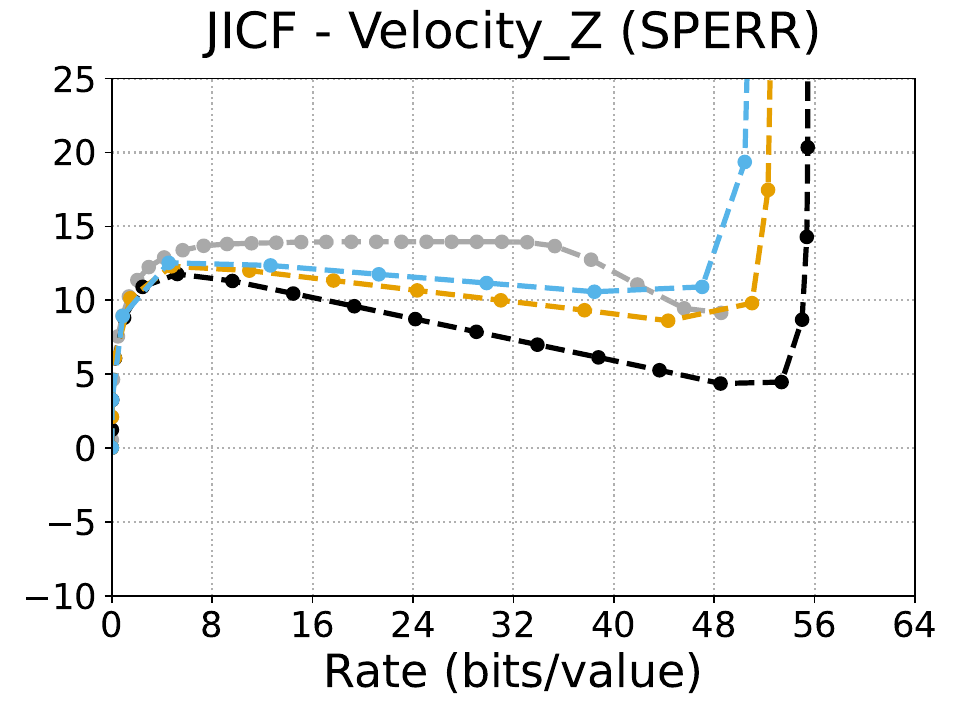}\hfill%
 \includegraphics[width=.25\textwidth]{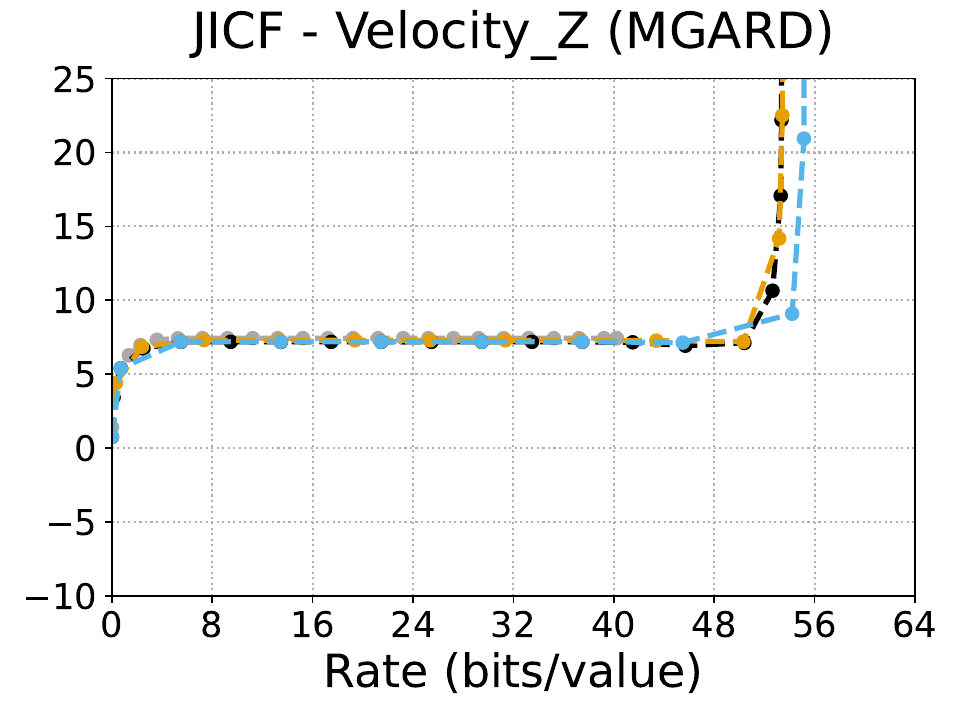}%
 \caption{%
   Accuracy gain (higher is better) vs.\ rate for
   multi-component approaches based on, from left to right, \zfp, \sz, \sperr,
   and \mgard.%
 }
 \label{fig:results2}
\end{figure*}

\subsubsection{Comparison with Nonprogressive Compressors}

For the remaining analysis, our chosen error metric is the recently proposed
\emph{accuracy gain}~\cite{multiposits,sperr},
\begin{equation}
  \gain = \log_2 \frac{\sigma}{\rmse} - \rate,
\end{equation}
where $\sigma$ is the standard deviation of the input data, \rmse is the root-mean-square
error, and \rate is the rate in compressed bits per scalar value.  \gain is related to
signal-to-noise-ratio (\snr, in $\text{dB}/(\text{bit}/\text{value})$) by $\gain =
\frac{\snr}{20 \log_{10} 2} - \rate \approx \frac{\snr}{6.02} - \rate$.

We plot $\gain(\rate)$, where a positive slope indicates that, at rate \rate, compression
is occurring (more bits of precision are consumed than are output); a negative slope
indicates expansion at that rate (more bits are output than are consumed).  A flat curve
implies that each bit emitted corresponds to one bit of precision consumed, as indicated
by a halving of the error, \rmse.  This type of information is difficult to infer in an
\snr plot. We further note that \gain, unlike \snr, accounts for both error and
rate---hence, an \gain value on its own is meaningful.  Except when approaching lossless
compression, where $\rmse \rightarrow 0$ and $\gain \rightarrow \infty$, \gain indicates
the cumulative number of bits per value that have been inferred by the compressor.

\begin{figure*}[tp]
 \centering%
 \includegraphics[width=.30\textwidth]{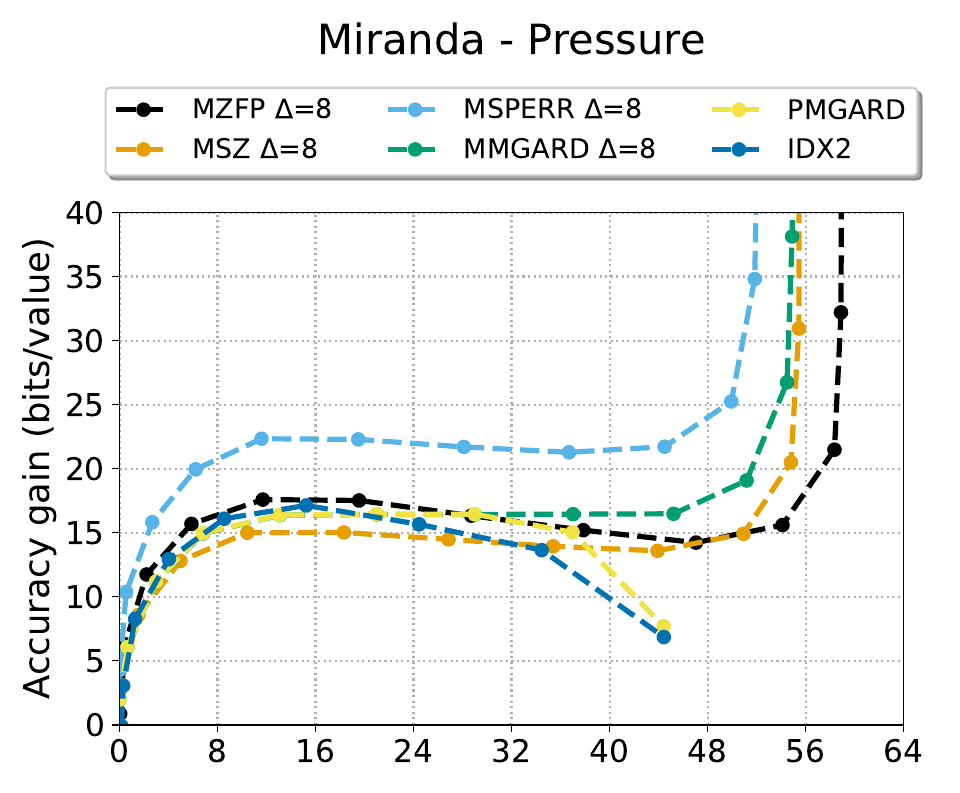}\ %
 \includegraphics[width=.30\textwidth]{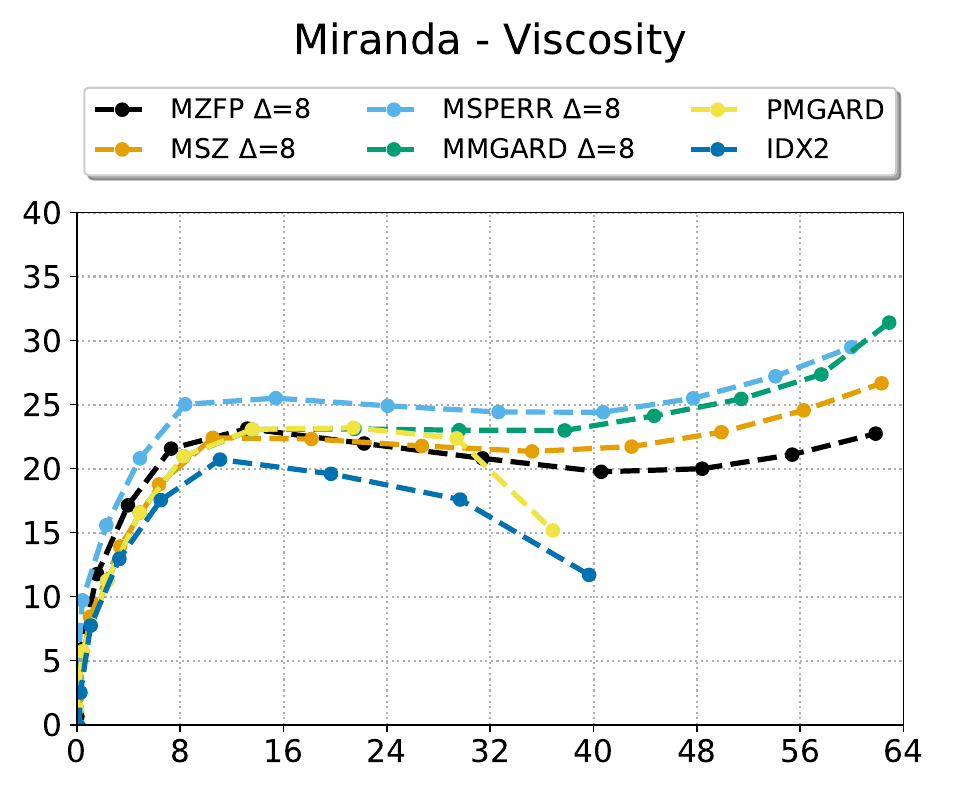}\ %
 \includegraphics[width=.30\textwidth]{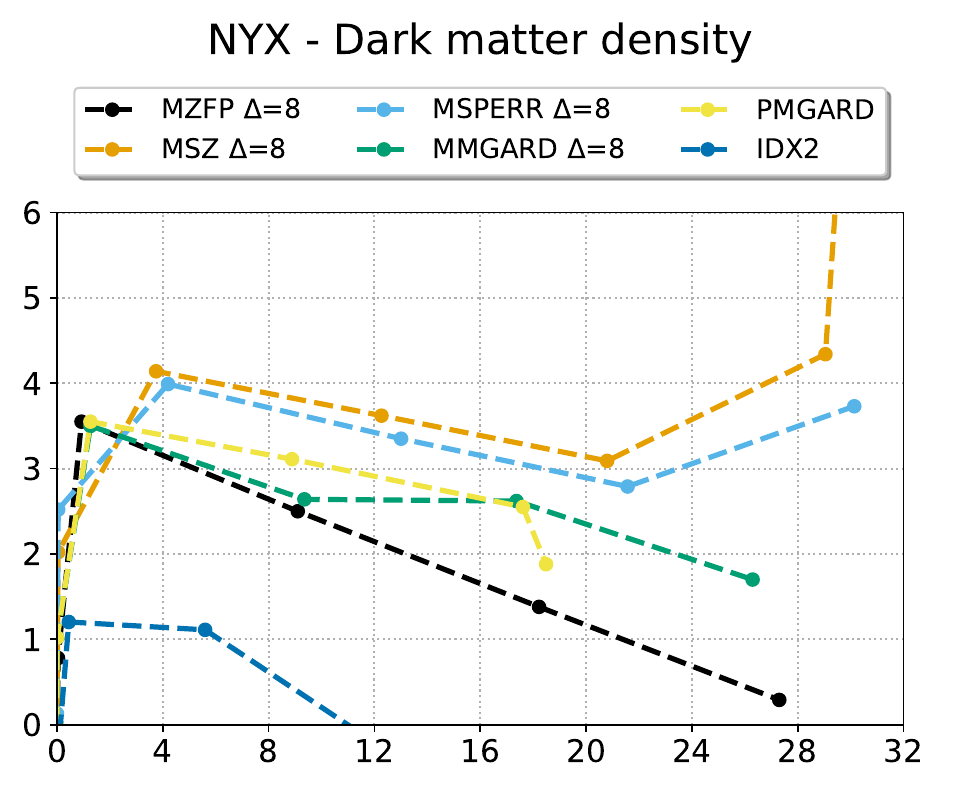}\\%
 \includegraphics[width=.30\textwidth]{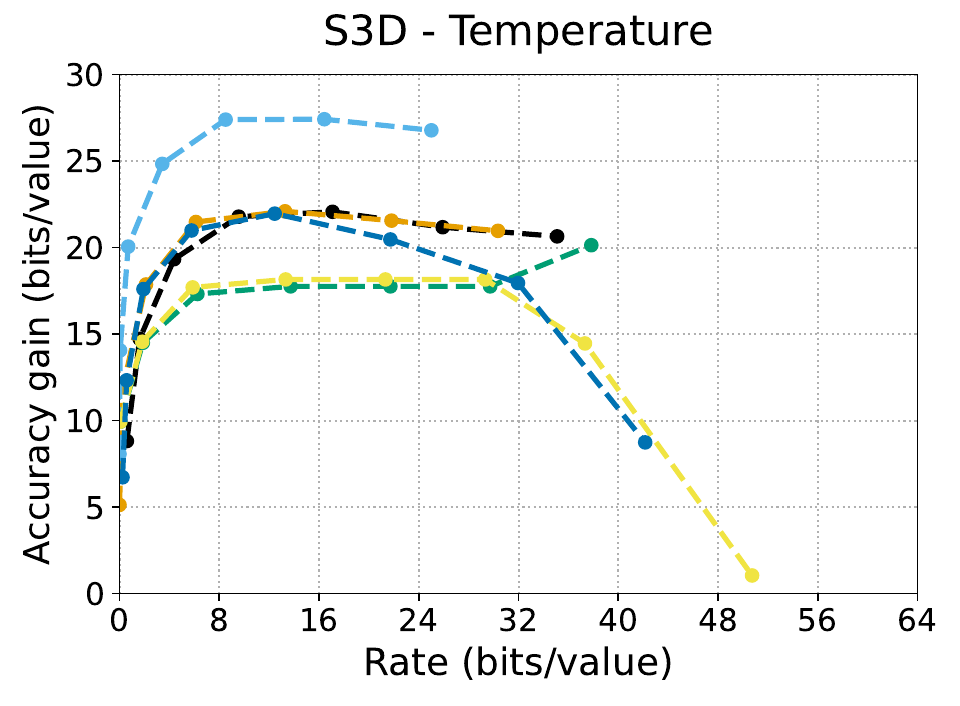}\ %
 \includegraphics[width=.30\textwidth]{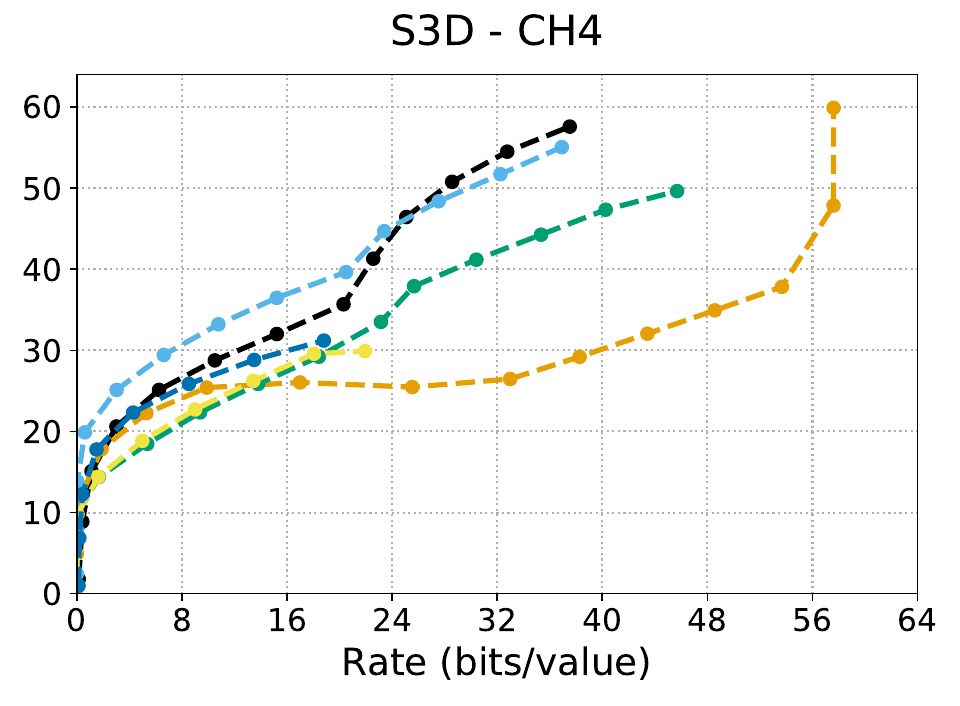}\ %
 \includegraphics[width=.30\textwidth]{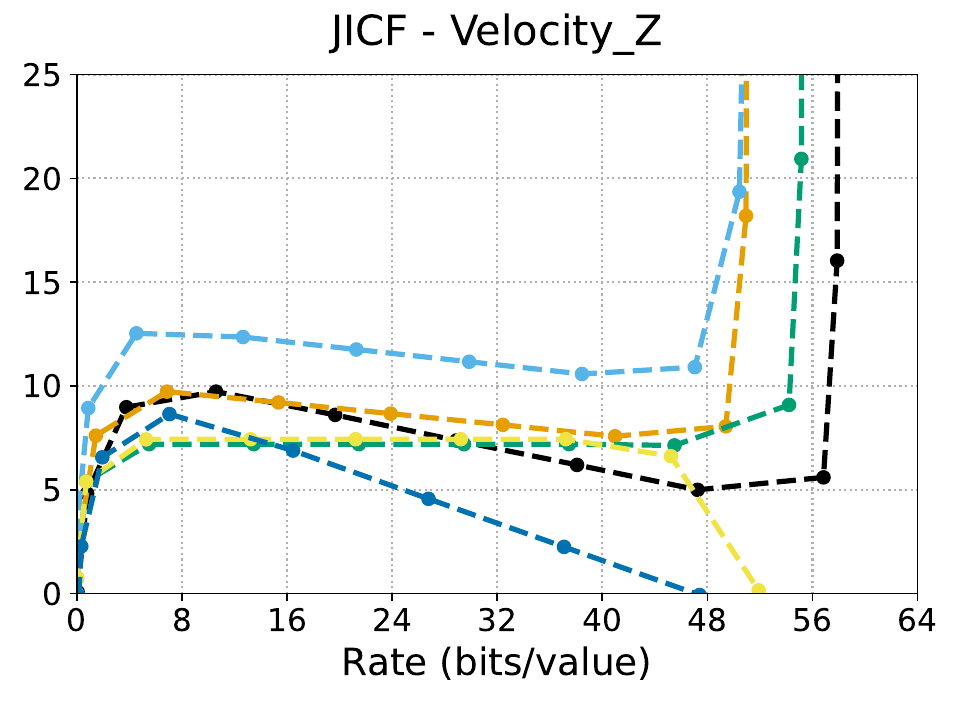}%
 \caption{%
    Accuracy gain (higher is better) comparison among our multi-component compressors
    and with \idxtwo and \pmgard.
 }
 \label{fig:progressive-comparisons}
\end{figure*}

\begin{figure}[t!]
  \centering%
  \includegraphics[width=0.9\linewidth]{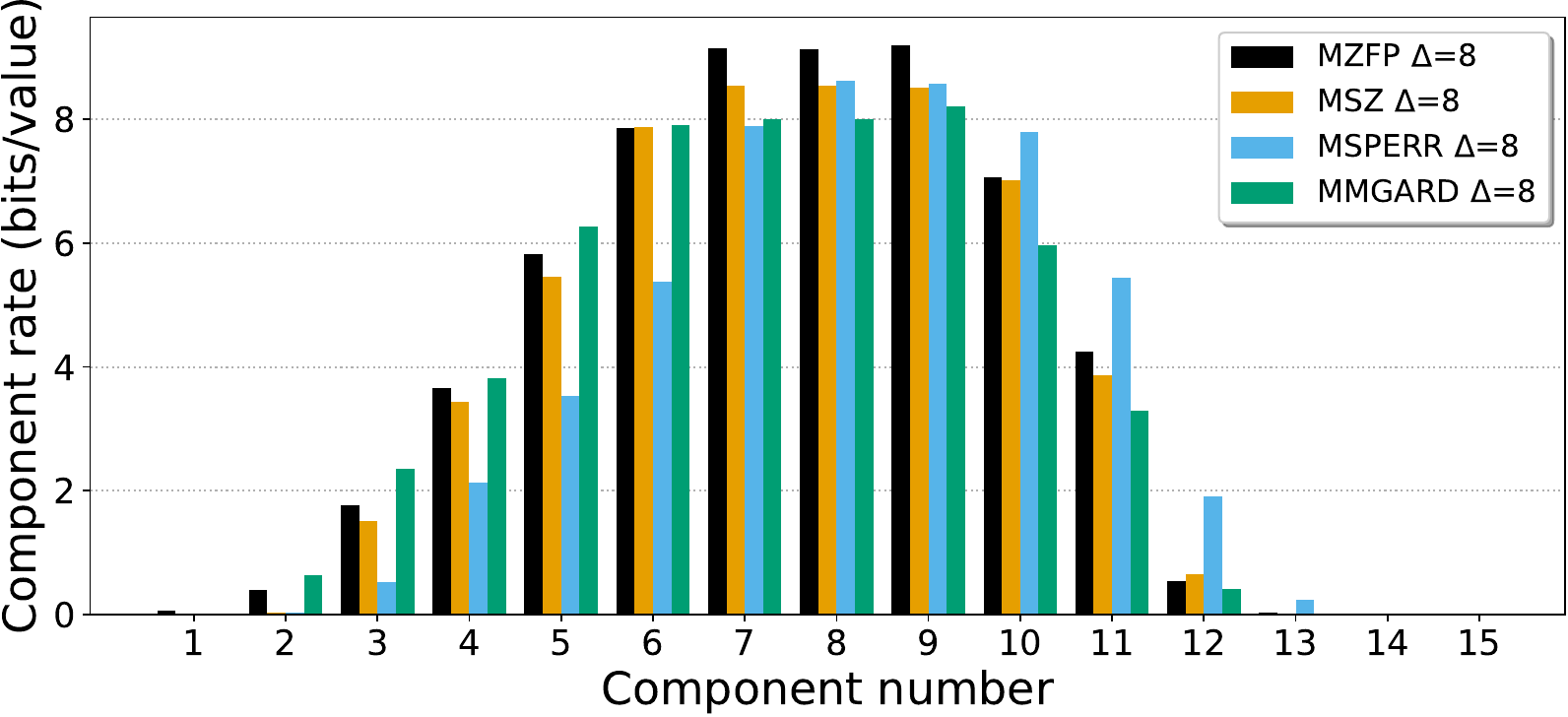}%
  \caption{
    Per-component rate, $\rate_i$, for the Miranda pressure
    field.  Components with
    $\rate_i > \gran = 8$ expand rather than compress due
    to high entropy.
  }
  \vspace{-2ex}%
  \label{fig:pressure-rates}
\end{figure}

\newpage
\Cref{fig:results2} shows accuracy gain plots for six data sets using
our four single- and multi-component compressors.  We note several general
trends.
First, the single-component compressors (gray curves) often---but not
always---do better, especially at low rates.
We point out, however, that these compressors, which do not support
progressive reconstruction, are re-run for each fixed tolerance, for
which they optimize their bit streams.  As discussed earlier, while
\zfp and \sperr utilize bit plane coding that in principle could
accommodate progressivity, the required changes to algorithms and
data structures are nontrivial.  We include these single-component
results here as gold standards for calibrating the multi-component
results.

Second, coarser granularity generally leads to higher accuracy gain for multi-component
compressors. This is not surprising as finer granularity implies a larger number of
components, each of which incurs some storage overhead. The amount of overhead is related
to the relative spacing between the three multi-component curves. Notably, \mgard appears
to exhibit very low overhead, as the three curves largely overlap.

Third, the multi-component curves generally approach vertical asymptotes at high rates, as
the error by construction must approach zero until we eventually achieve lossless
compression (see later discussion).  This unintended benefit of our approach makes it
possible to use existing lossy compressors for use cases like checkpoint-restart and data
dissemination, both of which often demand lossless compression.

Fourth, we often see the multi-component curves dip over mid-range rates, which can be
explained by increasing randomness in components to the point where the data expands
rather than compresses.  (One could, of course, counter this by storing the data in raw
form whenever expansion occurs.)  This expansion is evident from
\cref{fig:pressure-rates}, which plots the per-component bit rate, i.e., the contribution
of each component to the overall storage cost.  Whenever the per-component rate exceeds
the granularity, $\gran = 8$, in this plot, as occurs for components 7--9 that represent
some of the least significant, near-random bits in the field values, expansion rather than
compression occurs.  For later components (and thus higher cumulative rates in
\cref{fig:results2}), errors eventually reach zero for many field values, promoting
compression, often with very low per-component rates.

We also make some compressor-specific remarks.  As single-component compressors, \zfp and
\sperr, which both use bit plane coding, tend to reach stable plateaus in \gain at
mid-range rates, where less significant bits are essentially incompressible and emitted
verbatim.  \mgard largely follows this trend, as well, whereas the behavior of \sz is
quite different.  We conjecture that the conspicuous drop in \sz accuracy gain around
8~bits/value is due to its use of Huffman coding of high-precision integer correctors.
For small enough tolerances, the number of different corrector values increases to the
point where it approaches the number of data values, essentially precluding
compression---a phenomenon noted in~\cite{fpzip}.  In this case, the \sz storage overhead
of its Huffman table becomes substantial. This overhead remains small by compressing the
data as multiple components, with the tolerance for each component relatively large in
relation to its range (i.e., on the order $2^{-\gran}$).  Hence, multi-component \sz often
performs better.  Finally, the consistently low accuracy gains observed for Nyx are due to
poor spatial correlation (``smoothness''), with per-axis autocorrelation below 0.79,
compared to above 0.95 for all other data sets in our study.

\subsubsection{Comparison with Progressive Compressors}

We here compare our multi-component framework with two recent progressive compressors:
\idxtwo~\cite{Hoang2021} and \pmgard~\cite{LiGoCh21} (aka.\ \mdr).  We note that these two compressors
support progression in both precision and resolution; to allow an apples-to-apples
comparison, we enforce full resolution.  \Cref{fig:progressive-comparisons} plots
$\gain(\rate)$ for these two and our four progressive compressors using the same data sets
as in \cref{fig:results2}.  We set progressive granularity to $\gran = 8$, at which we
evaluate all compressors. As is evident, at least one of our multi-component compressors
does as well as both \idxtwo and \pmgard, and often significantly better.  Showing all
compressors in the same plot reveals that \msperr generally performs the best on average.
Furthermore, whereas accuracy gains for \idxtwo and \pmgard drop at high rates, our
compressors all trend upward and eventually achieve lossless compression.

\begin{figure}[tp]
  \centering%
  \hspace{0.003\textwidth}%
  \includegraphics[width=0.924\linewidth]{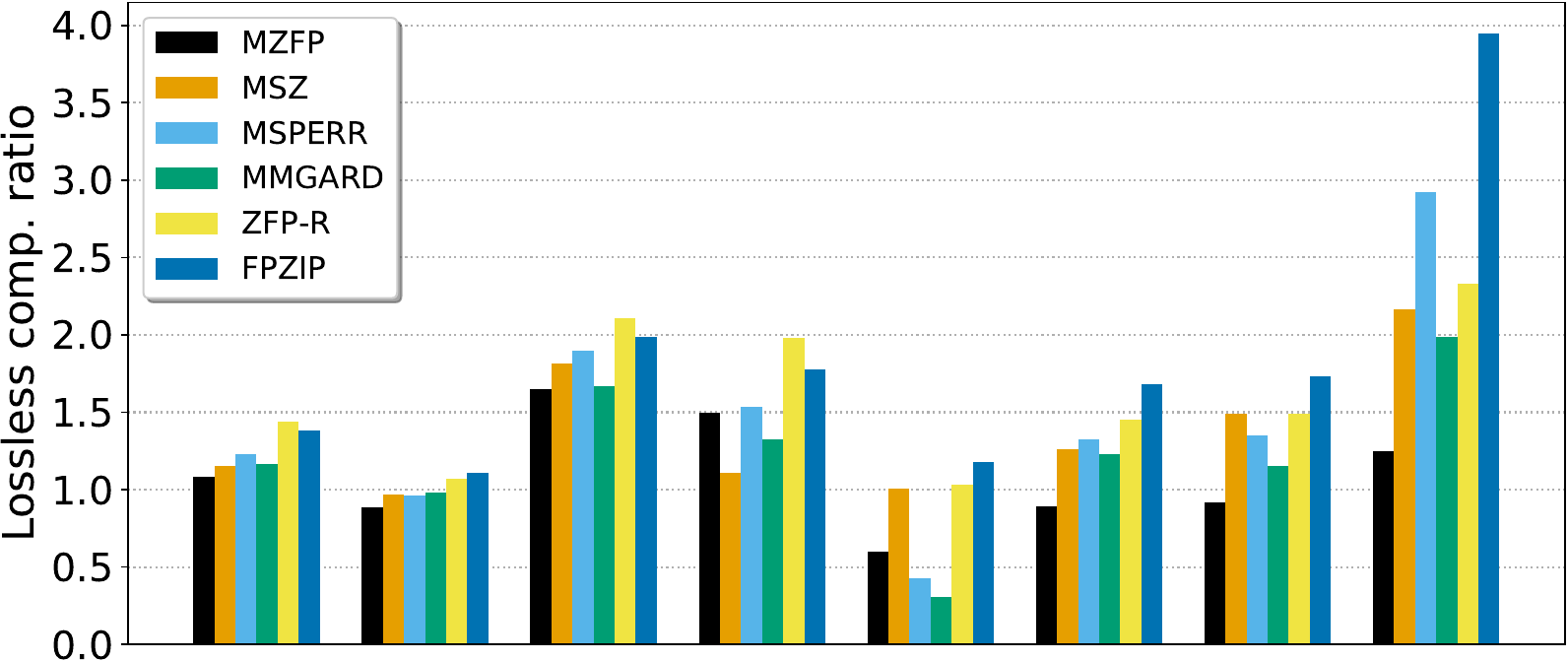}\\%
  \includegraphics[width=0.93\linewidth]{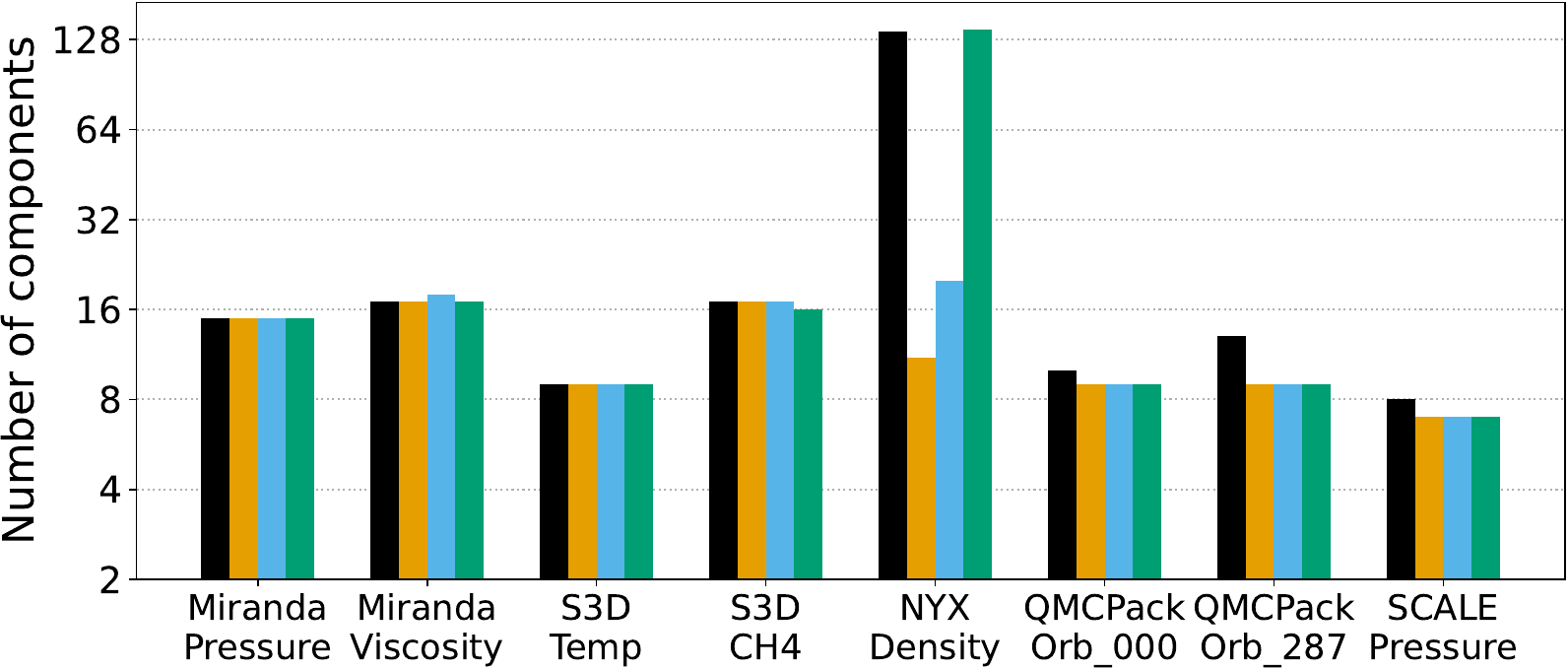}%
  \caption{%
    Lossless compression ratio (top) and number of components required by
    the multi-component compressors (bottom).
  }
  \label{fig:lossless}
\end{figure}

\subsubsection{Lossless Compression}

As alluded to, since each additional component's inclusion reduces error and since
precision is finite, we can achieve fully lossless compression using a sufficient number
of components. This allows using lossy compressors without additional redesign for use
cases that demand bit-for-bit exact reproducibility.

How competitive is such an approach in terms of compression ratio?  \Cref{fig:lossless}
plots lossless compression ratios for several fields.  For calibration, we also include
here results for compressors designed to be lossless: \zfpr (the ``reversible'' mode of
\zfp) and \fpzip.  While our multi-component compressors sacrifice some compression, they
still give reasonable compression ratios and often fare not much worse than the
specialized methods.  Note that \idxtwo and \pmgard do not allow for such lossless
compression.

\begin{figure}[tp]
  \includegraphics[width=\linewidth]{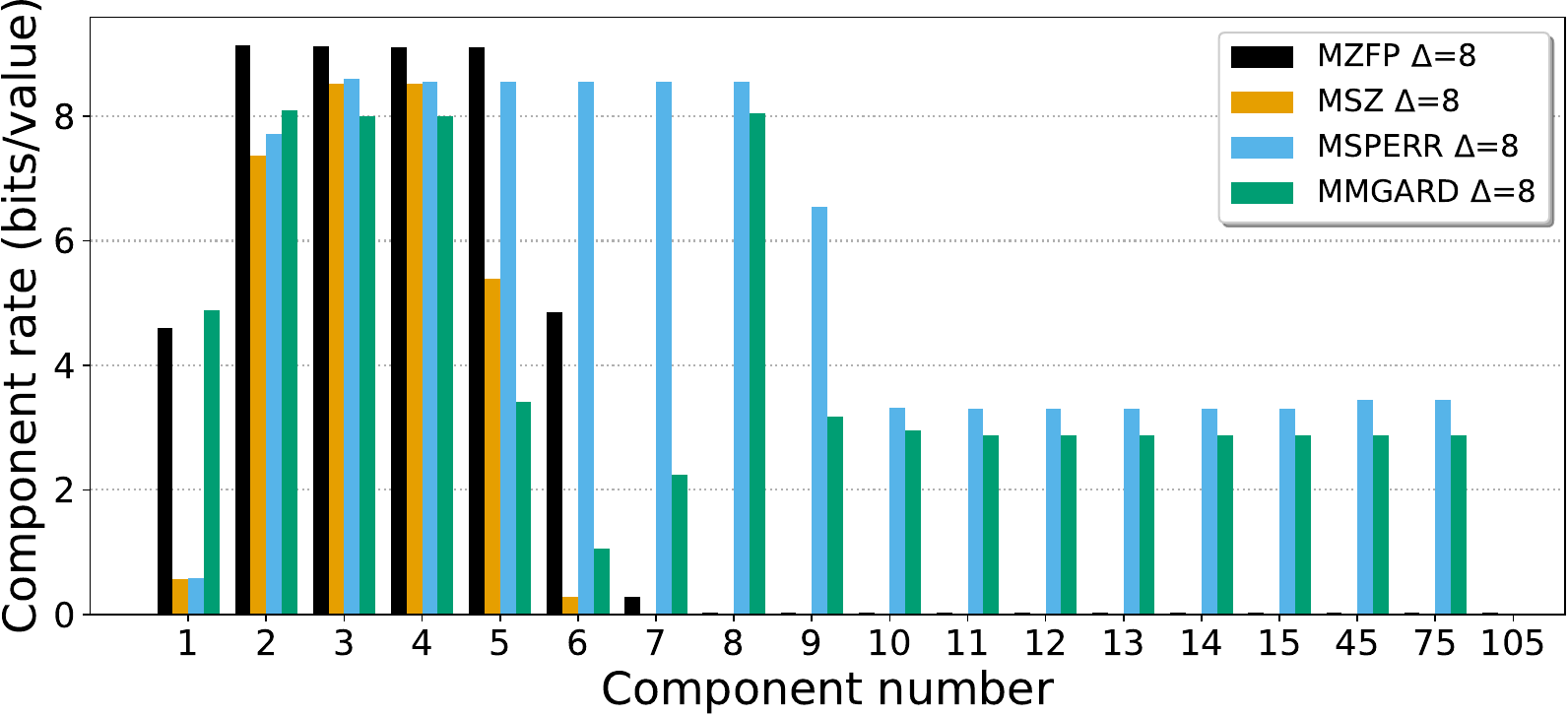}%
  \caption{%
    \msperr and \mmgard sometimes require many components to achieve
    lossless compression for data containing zeros (here the Nyx
    dark matter density field).
  }
  \label{fig:lossless-nyx}
\end{figure}

We highlight one challenge that \msperr and \mmgard have with lossless compression of data
containing zeros.  Because these compressors perform floating-point arithmetic involving
basis functions with wide support, many zero-values are often initially approximated as
slightly nonzero.  With each subsequent component, those nonzeros are brought closer to
zero, but getting all the way to zero requires going through tolerances spanning the
subnormal range, with associated components that each have a nonnegligible coding cost.
This phenomenon is illustrated in \cref{fig:lossless-nyx}.

\vspace{0.5ex} 
\subsection{Performance Analysis}
\label{subsec:results_performance}

We now investigate the serial execution time of \cref{alg:multicomp1,alg:multicomp2} for
constructing the multi-component representation and later reconstructing a field,
respectively.  Since the overall trends of the four multi-component approaches are
consistent across different data sets, we focus only on the results for the Miranda
pressure field.  We refer the reader to the supplemental material for other fields.  For
the multi-component and progressive compressors, the timings reported are cumulative with
respect to the number of components, while in the case of the single-component
compressors, we report the time to (de)compress the whole data set (as a single
component).

\subsubsection{Multi-Component Construction and Reconstruction}
\label{subsubsec:times}

\begin{figure*}[!tb]
 \includegraphics[width=.25\textwidth]{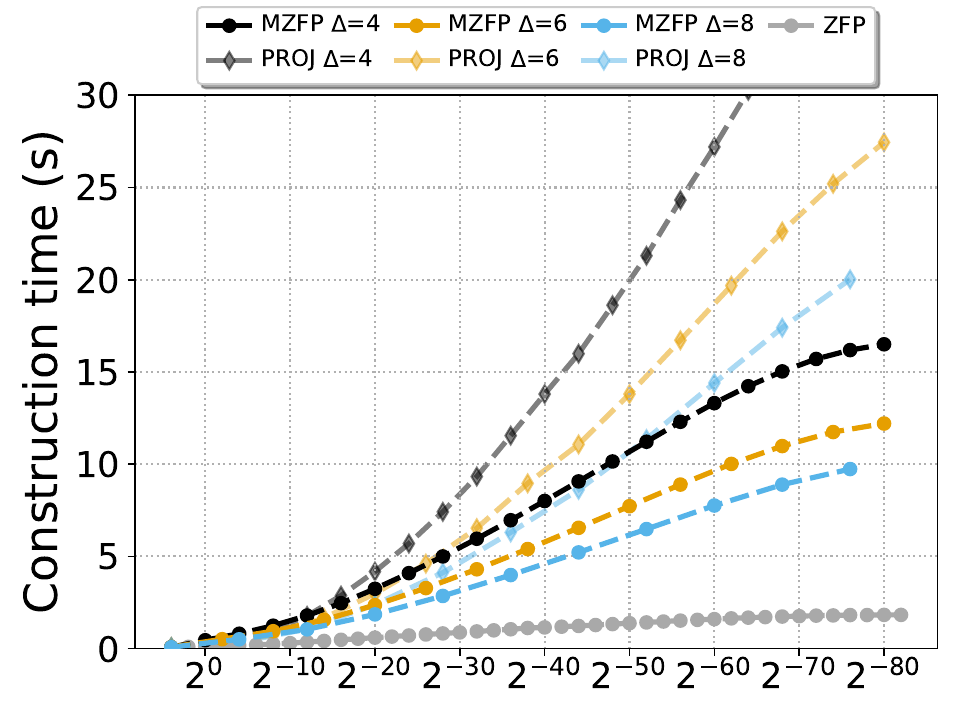}%
 \includegraphics[width=.25\textwidth]{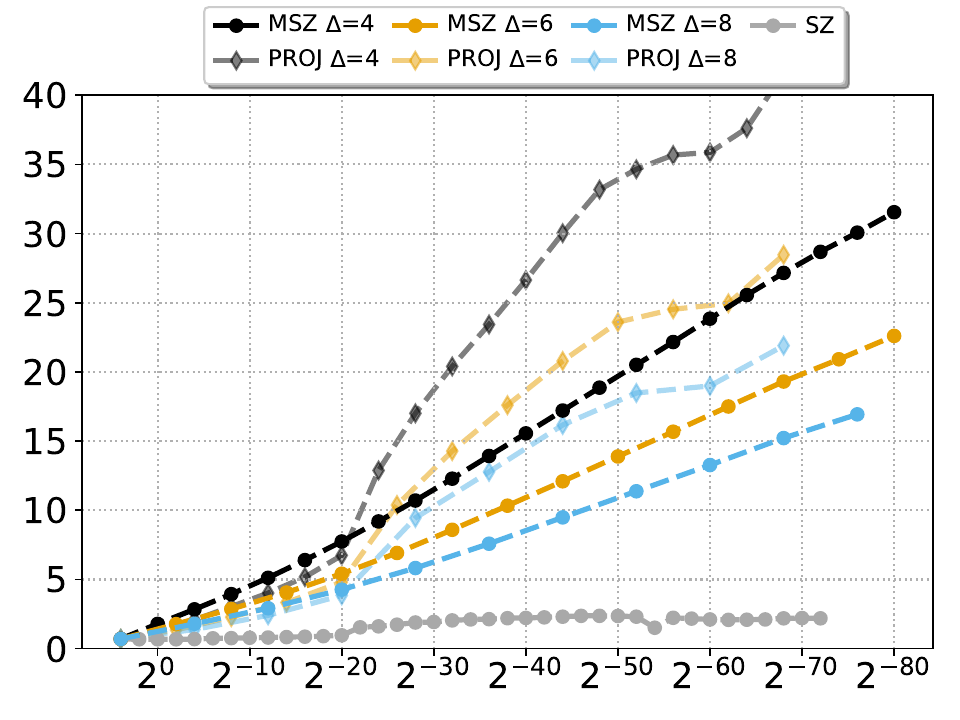}%
 \includegraphics[width=.25\textwidth]{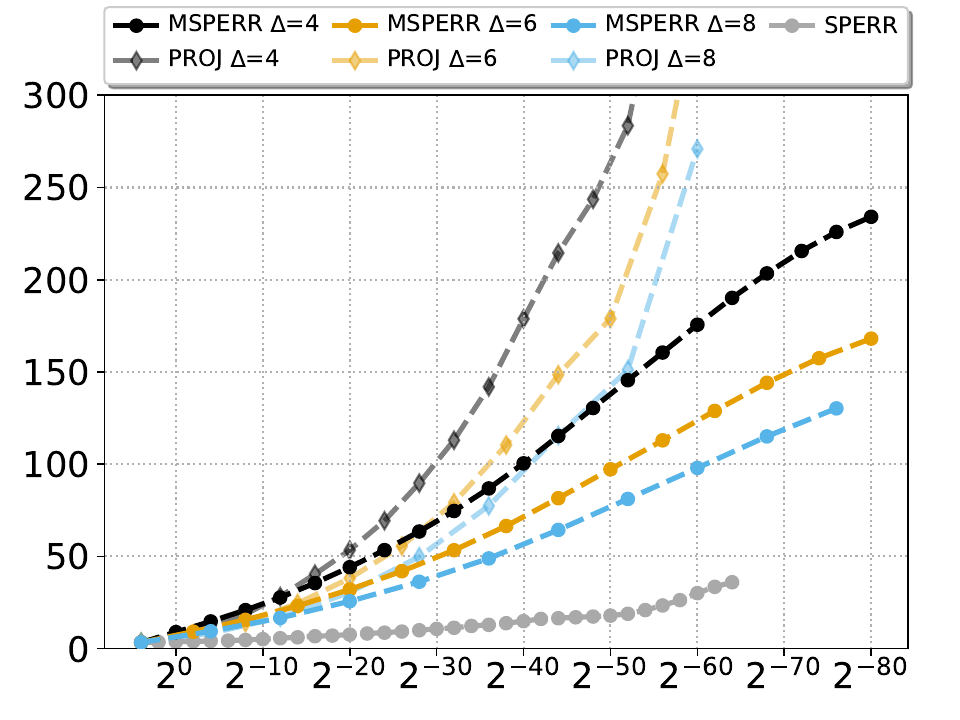}%
 \includegraphics[width=.25\textwidth]{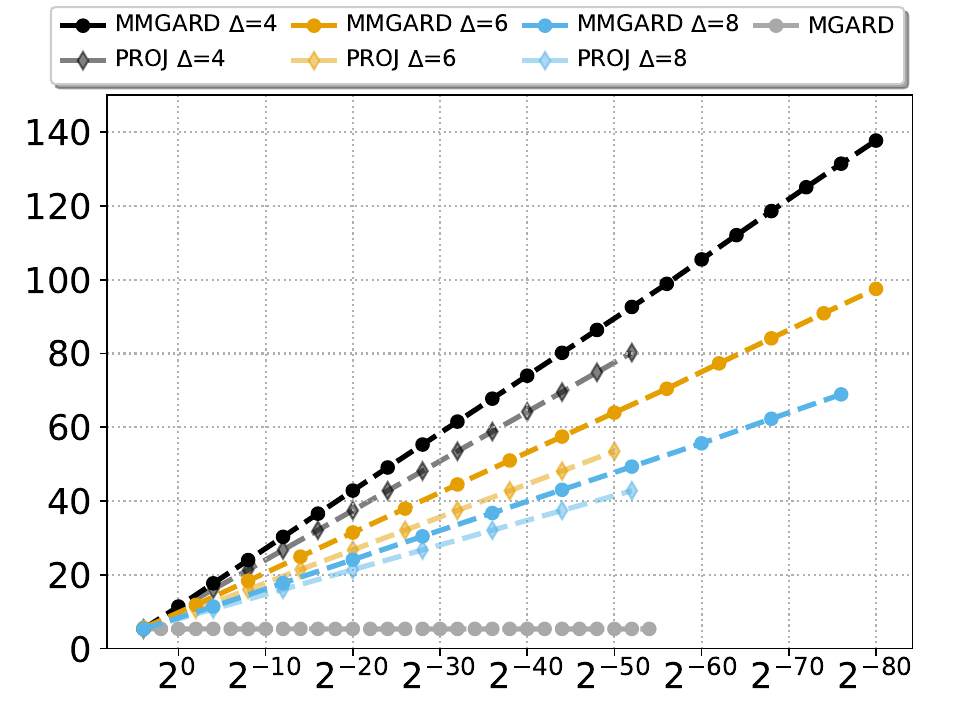}\\
 \includegraphics[width=.25\textwidth]{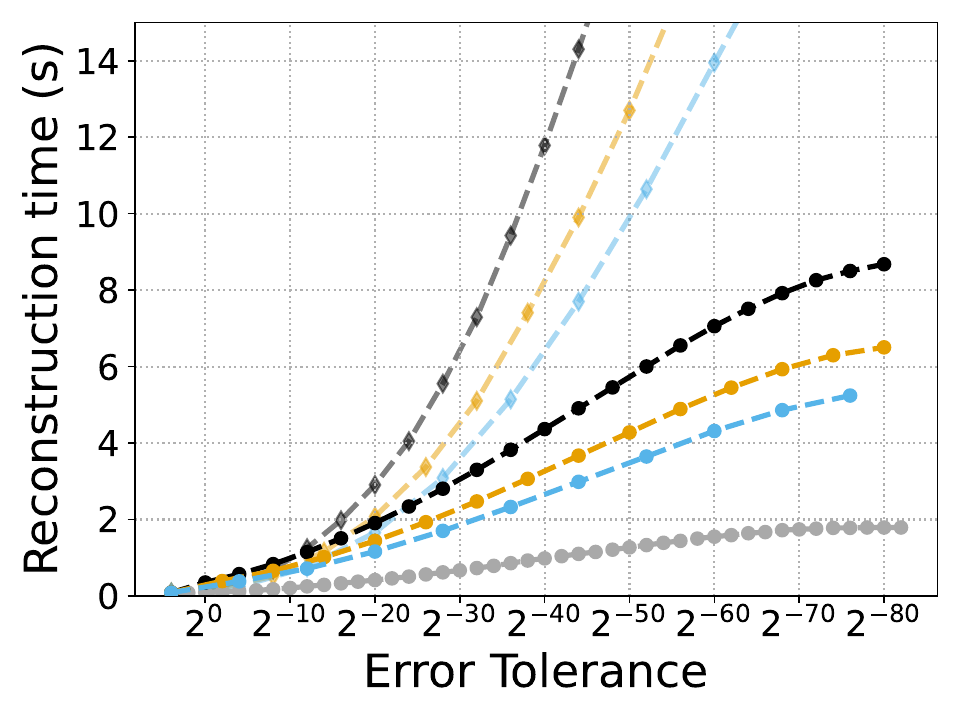}%
 \includegraphics[width=.25\textwidth]{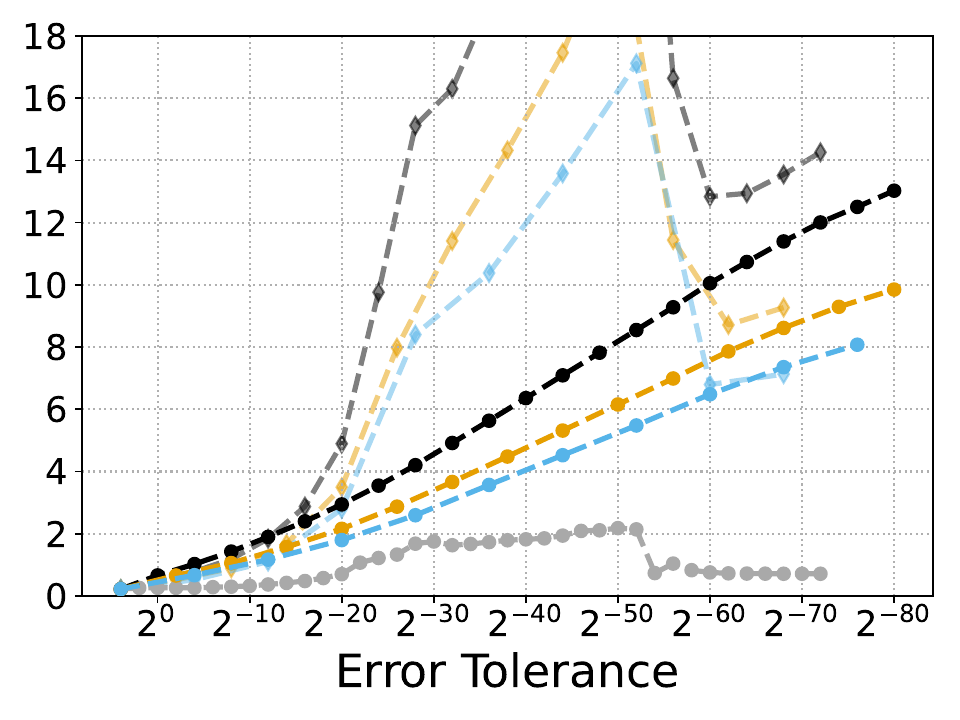}%
 \includegraphics[width=.25\textwidth]{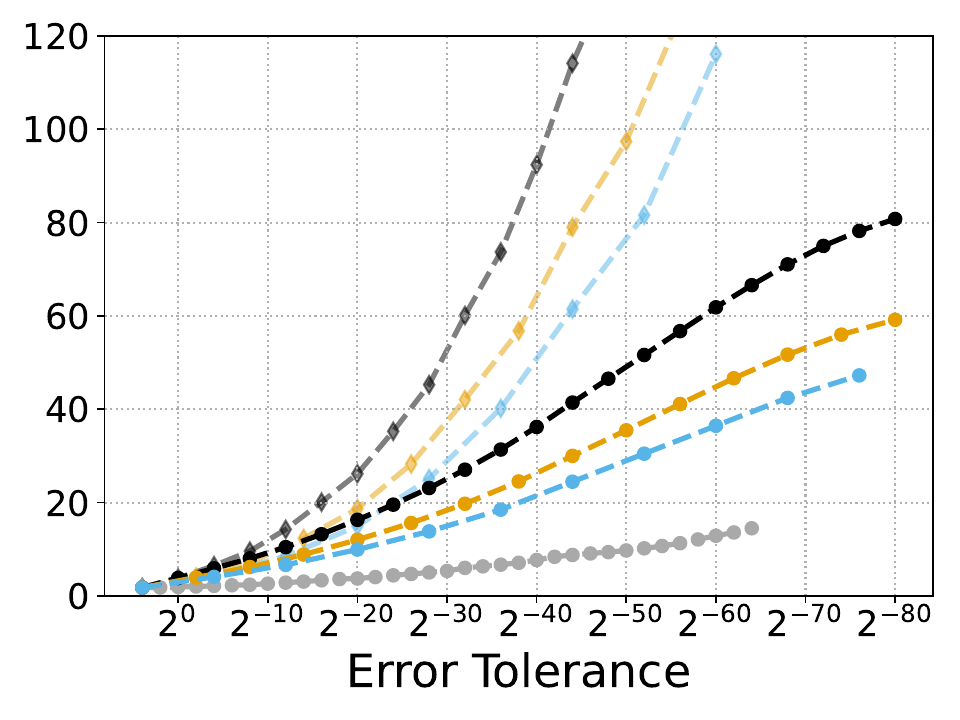}%
 \includegraphics[width=.25\textwidth]{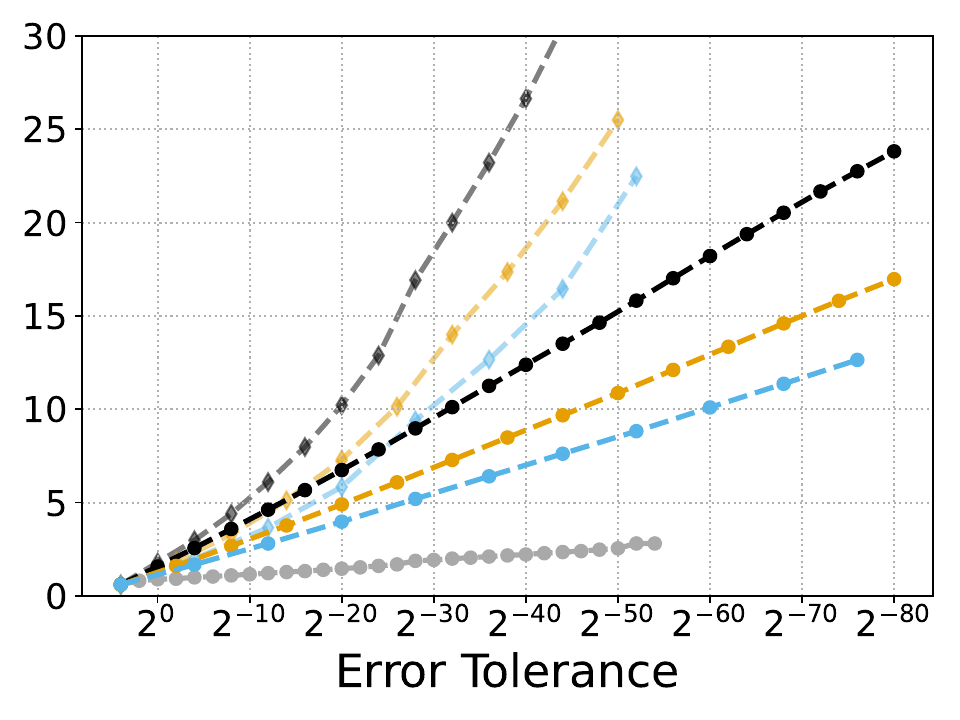}%
 \vspace{-0.5ex}%
 \caption{%
   Construction time (top) and reconstruction time (bottom) vs.\ error tolerance
   for our multi-component approach based on (from left to right)
   \zfp, \sz, \sperr, and \mgard. The multi-component approaches exhibit longer
   execution times than their single-component counterparts, though generally
   less time than linearly scaling the single-component time by the corresponding
   number of components (labeled \proj).
 }
 \label{fig:results7}
 \vspace{-0.5ex}%
\end{figure*}

\Cref{fig:results7} plots construction (top) and reconstruction (bottom) time, i.e.,
execution time for \cref{alg:multicomp1,alg:multicomp2}, for the multi-component
compressors based on three granularities as a function of the smallest error
tolerance $\tol_n$ requested by the user. For comparison purposes, we include the
execution time of each corresponding single-component compressor and its projected time,
denoted \proj, when executed as many times as iterations made by the corresponding
multi-component compressors. Analyzing first the top four plots, we note
that, except for \mgard,%
\footnote{We found that the \mgard compression time is, counterintuitively, independent
of tolerance and bit rate; this behavior needs further investigation.}
all multi-component compressors show execution times
in the interval between the single-component compressor and projected times. In addition,
the gap between single- and multi-component compressor time decreases with coarser
granularity, \gran.
This happens because, for a given error tolerance, fewer components
are generated with larger \gran, thus the number of calls to the most time consuming steps
in \cref{alg:multicomp1} (Lines~$4$ and~$5$) is also smaller.  Looking at the slope
of the curves, we note that \mzfp, \msz, and \msperr start with small inclinations,
transition to larger ones as new components get added, and tend to finish with a flat
profile when approaching lossless compression. This behavior is related to the inverted
U-shape of the component rates shown in \cref{fig:pressure-rates}, since compression time
tends to be proportional to compressed 
size.
Note that in practice, one would set a single finest error tolerance during
construction instead of repeating compression runs with different tolerances, as done
for these plots.

Moving our discussion to reconstruction time, we note that \mzfp, \msz and
\msperr behave very similarly to the case of construction, and the same conclusions drawn
above apply here. The main difference between construction and
reconstruction time is that, for a given error tolerance or number of components, the
latter is always smaller for any granularity of the multi-component approach. This is
expected since the main cost of \cref{alg:multicomp2} is running the decompressor, a task
also included in \cref{alg:multicomp1}. Lastly, we note that \mmgard shows behavior
different from before. In fact, the decompression algorithm of \mgard depends on the
amount of data decoded, and since this varies with the number of components, as shown in
\cref{fig:pressure-rates}, the reconstruction time for \mmgard becomes bounded by the
projected worst-case time.

\subsubsection{Comparison with Progressive Compressors}
\label{subsubsec:progressive_performance}

\begin{figure}[!tb]
\includegraphics[height=1.55in]{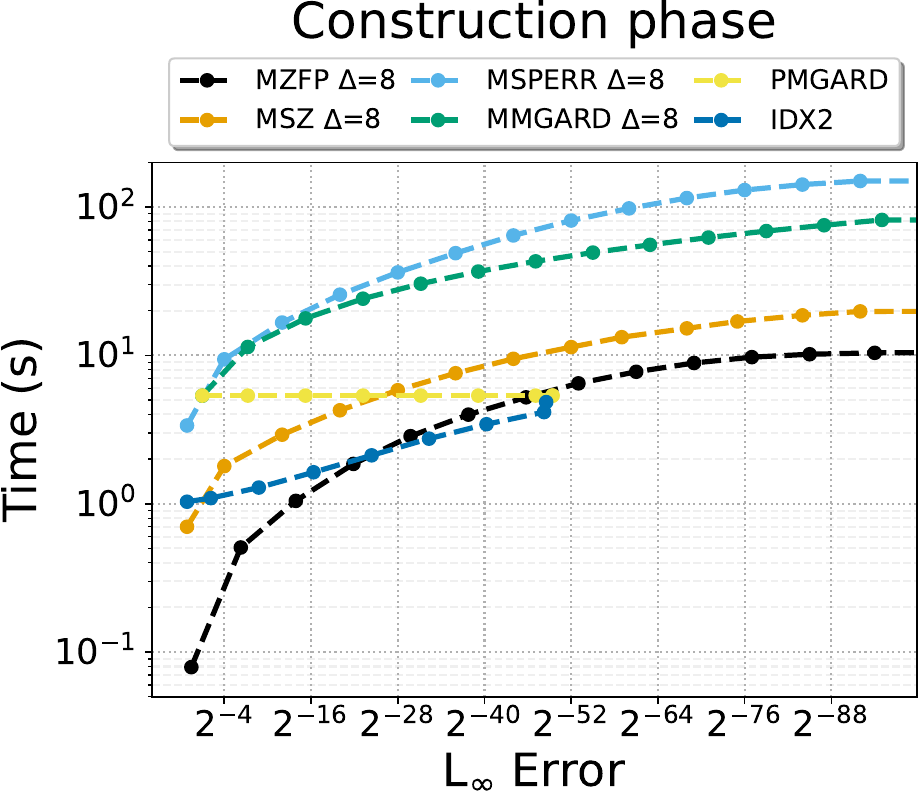}%
\hfill%
\includegraphics[height=1.55in]{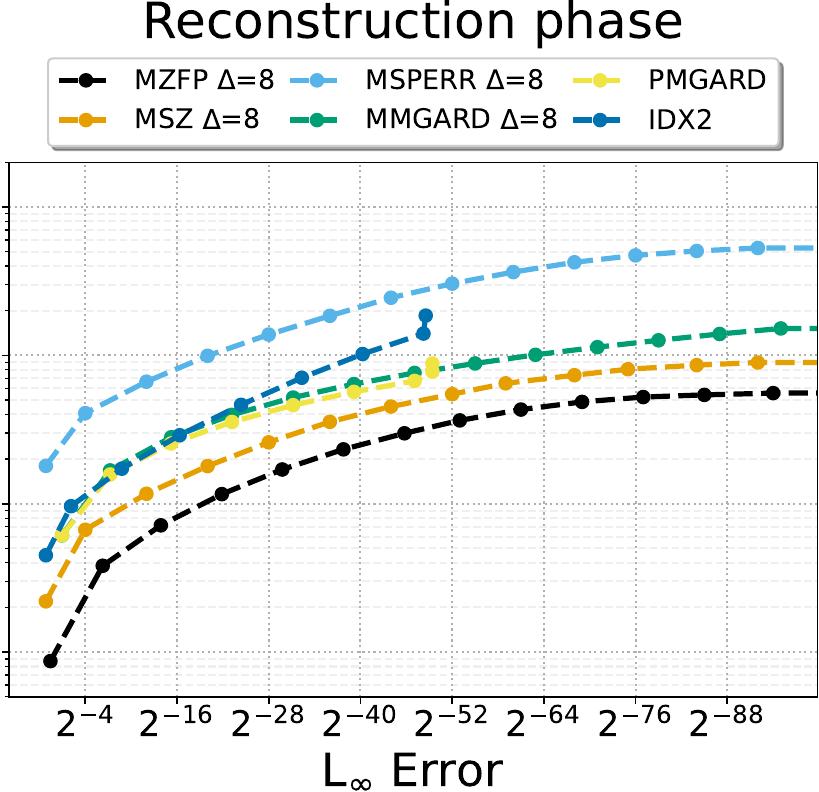}%
\vspace{-0.5ex}
 \caption{Construction (left) and reconstruction (right) time vs.\ error
   tolerance for the multi-component and progressive compressors. Performance
   wise, our multi-component methods, especially \mzfp and \msz, are competitive
   with if not superior to \idxtwo and \pmgard.}
 \label{fig:results8}
\vspace{-1.5ex}
\end{figure}

Lastly, in \cref{fig:results8}, we evaluate how the performance of the
multi-component techniques compares against \idxtwo and \pmgard. For this purpose, we
select the results for $\gran = 8$.
We note that \idxtwo does not support an absolute error tolerance; rather,
it is driven by a user-specified target $L_2$ error.  To allow a fair comparison,
we therefore plot (de)compression time as a function of \emph{measured} maximum
absolute ($L_\infty$) error, which strongly correlates with error tolerance
for the other compressors.

Looking first at compression times, we note that the
progressive compressors show a considerable overhead starting from the loosest error
tolerance. This is associated with the additional metadata required by them to achieve
progressivity. Next, we observe that the multi-component compressors \msperr and \mmgard
are less efficient than the progressive compressors mainly, and this is due to the higher
costs of running their underlying compressors (\sperr and \mgard, respectively). On the
other hand, \mzfp and \msz benefit from their faster underlying compressors. As a
result, these techniques are similar in performance to \idxtwo and \pmgard.

Finally, moving to data reconstruction times, all multi-component techniques but \msperr
are faster than \idxtwo, while \mmgard and \pmgard give similar performance.
From these results, we can affirm that our multi-component approach is competitive in
performance with other progressive compressors available in the literature.

\subsection{Progressive Use Case}
\label{subsec:use_case}

We conclude this section with an example use case of progressive compression that
highlights varying precision requirements for different visualization tasks involving
derivatives.  Spatial derivatives are central to numerous scientific visualization and
data analysis techniques, including gradients for volume rendering~\cite{Correa11}, curl
for vortex detection~\cite{Gunther18}, and second derivatives for extraction of ridges and
Lagrangian coherent structures~\cite{Shadden05}.  As is well
known~\cite[\S{5.1.2}]{Sauer11}, differential operators magnify round-off and, by
extension, compression errors---the round-off error in the $i^\text{th}$ derivative
follows $O(h^{-i})$, where $h$ is the grid spacing.  Thus, as $h$ approaches zero,
derivative estimates are increasingly susceptible to compression artifacts.

\begin{figure*}[tp]
\includegraphics[width=\linewidth]{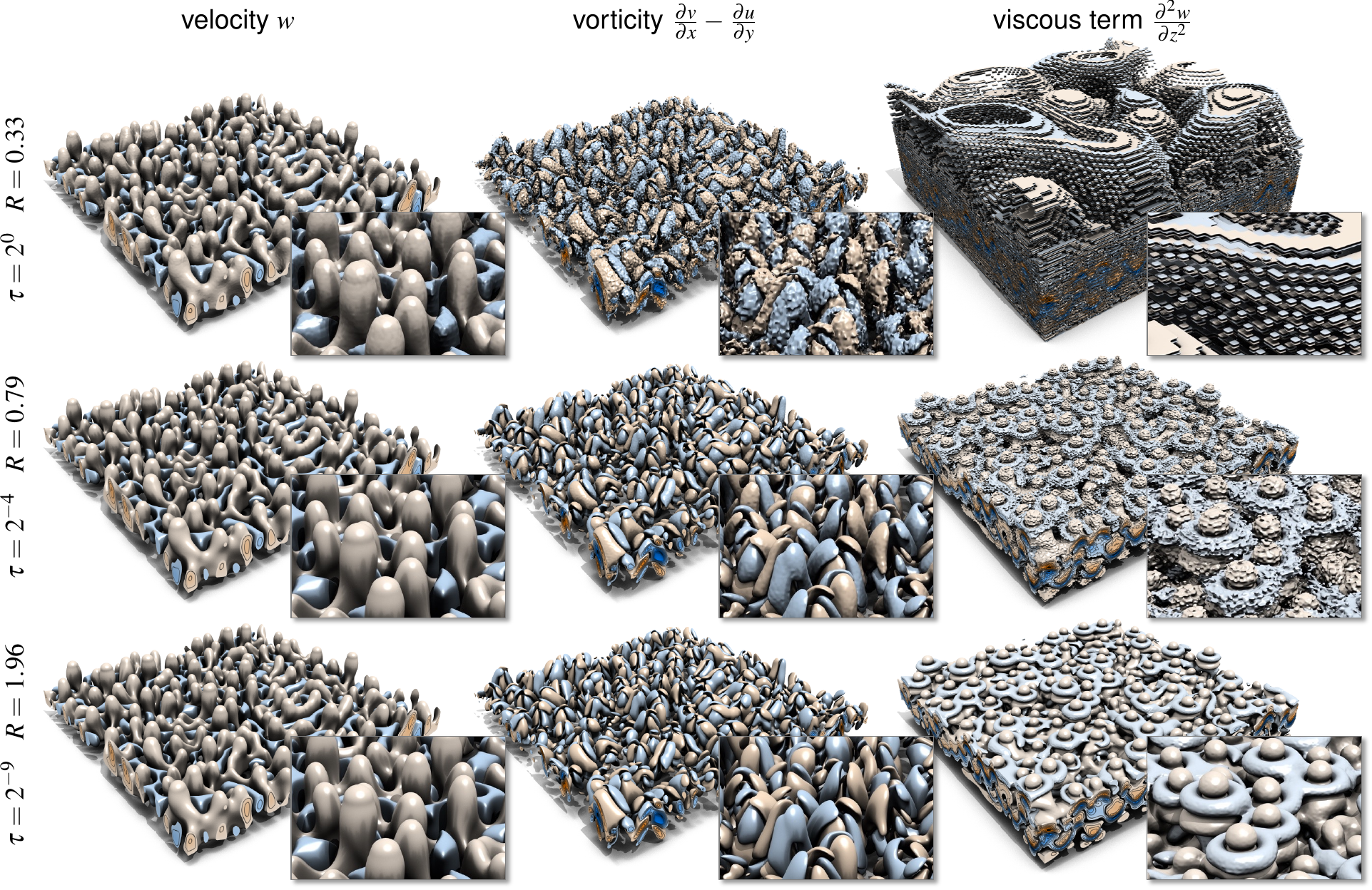}
\caption{%
Magnification of \mzfp compression errors by differential operators of increasing order
(left to right), requiring progressively smaller error tolerances, \tol, and
correspondingly higher cumulative bit rates, \rate, via additional components (top to
bottom).
}
\label{fig:results9}
\end{figure*}

In \cref{fig:results9}, we visualize the third component of the Miranda velocity field,
$\mathbf{u} = (u, v, w)$, which at even a relatively high tolerance $\tol_1 = 1$ yields a
quite acceptable result (top left panel).  The middle column shows the third component
$v_x - u_y$ of the vorticity field, $\nabla \times \mathbf{u}$, computed from the
progressive reconstruction, which at the largest tolerance exhibits significant artifacts.
This is because the finite-difference operator used to compute the curl greatly magnifies
the compression errors in $\mathbf{u}$.  At a tighter tolerance $\tol_2 = 2^{-4}$,
however, the vorticity is relatively artifact free (second row).  The rightmost column
shows the second spatial derivative, $w_{zz}$, a quantity related to viscosity in viscous
flows.  Clearly $\tol_1$ is insufficient, and even $\tol_2$ results in major artifacts.
The higher sensitivity of the second derivative operator requires an even finer tolerance,
$\tol_3 = 2^{-9}$, for acceptable results (bottom row).  These varying degrees of
sensitivity of data analysis and visualization tasks to compression errors is why it is
difficult to a priori choose a single error tolerance and why progressive compression is
an attractive solution.

For additional examples of progressive visualizations and quantitative results, we direct
the reader to the extensive supplemental material accompanying this paper.

%% file: conclusion.tex
\section{Discussion and Conclusion}
\label{sec:conclusion}

We have developed a framework for progressive-precision compression and reconstruction of
floating-point scalar fields based on the notion of a multi-component expansion, in which
independently-compressed components are successively added to form an ever more faithful
approximation.  Extensive experiments with real data show that our approach yields
competitive rate-distortion tradeoff and performance with two state-of-the-art progressive
compressors, while being far 
prior work, we support fully lossy to lossless compression while requiring no 
specializations to ensure losslessness.  As presented, our framework supports a 
sequence of absolute error bounds specified by the user at any desired granularity.

\subsection{Limitations}
We would like to acknowledge some limitations of our approach.  First,
storage of multiple components may introduce per-component overhead
paid only once in conventional single-component compression, and
which would likely be absent if one painstakingly re-engineered an
otherwise capable single-component compressor to support progressive
access, like the two compressors we compare with.
The per-component overhead is felt also in compression and decompression
time, though our framework achieves competitive and even superior
performance over state of the art.
Any such overhead is best hidden using fewer components, which
unfortunately implies less fine-grained access.
We do note, however, that GPU implementations of \zfp~\cite{cuzfp},
\sz~\cite{cusz}, and \mgard~\cite{mgardx},
which could easily be used within our framework, achieve
throughputs that are two to three orders of magnitude
higher than per-node I/O bandwidth to high-end parallel
file systems, effectively eliminating the performance cost of
having to compress multiple components.

Contrary to representations designed for in-memory compressed
storage~\cite{Ning92,Schneider03,Fout07,Li14,Guthe16}, our framework, like other
progressive compression techniques~\cite{Hoang2019,Hoang2021,Li2019,LiGoCh21,amm},
incrementally refines an uncompressed representation, \vapprox.  Enough CPU or GPU memory
for this reconstruction as well as a temporary buffer for one decompressed component,
\vcomp{i}, is assumed.  The space for this temporary buffer can be significantly reduced
using data chunking techniques, as employed for example in \tthresh~\cite{tthresh},
\sperr~\cite{sperr}, and---to a more extreme degree---\zfp, where single $d$-dimensional
blocks of $4^d$ scalars can be decompressed at a time and added into \vapprox.

We may also reduce the in-memory storage cost for \vapprox.  Though specific to \mzfp, we
may keep the components $\{\vzcomp{i}\}$ in compressed form and decompress and add blocks
on demand to form requested pieces of \vapprox.  As we have seen, the overhead of multiple
compressed components over a single compressed sum of components is often marginal.  We
also suggest a more general hybrid approach that maintains \vapprox not in
double-precision floating point but as a \zfp array.  Toward this end, \zfp~1.0.0 supports
both fixed-rate and error-bounded variable-rate arrays, allowing significant storage to be
saved while retaining fast random access and sequential updates.  As we are unaware of any
solution that currently supports progressive updates of in-memory compressed arrays, we
view such an approach as an exciting new research opportunity.

While we are not married to any particular error metric or
parameters that drive the compression process, supporting
pointwise \emph{relative} error bounds may be challenging.
One could, however, evaluate relative errors a~posteriori and store
such per-component errors as metadata.
Finally, our support for lossless compression is limited to
numerical data---NaNs, infinities, and signed zeros are not
handled.

\subsection{Benefits}
Our approach also has several distinct benefits.  For one, it is compressor
agnostic and can hence trivially be made to work with any lossy
compressor (and compressor parameters) or novel number representation,
like \posits~\cite{posits}.
In fact, one may even employ different compressors for different
components, e.g., by using computationally expensive but effective
compressors only for the first few components.
As we have seen, decomposing the data into multiple components can
in fact \emph{boost} compression (as in the case of \sz) or
accuracy (as in the case of \zfp and \mgard).
No custom progressive file format is needed; indeed, our approach
integrates directly into popular I/O formats/libraries like HDF5 and
ADIOS, with the sole change being how consumers interpret the additional
``component dimension.''
To meet any requested error tolerance, the prescribed tolerances or
measured errors would be stored as per-component metadata that the
user can consult to request a sufficient number of components.
To progressively increase precision, one simply decompresses
additional components and adds them to the current approximation,
with no need to maintain decompressor state.

One potential benefit of the complete decoupling of components is that they need not be
stored together.  As suggested in~\cite{LiGoCh21}, one may utilize slower media like tape
for the least significant, rarely accessed components.  One may take this one step further
and distribute and cache compressed components throughout the memory hierarchy, from CPU
and GPU RAM to node-local storage, parallel spinning disk, tape archives, and even remote
compute facilities and databases.  The distinct separation of components makes this
approach conceptually simple.

Finally, we acknowledge that our approach is embarrassingly straightforward.  However, we
view its simplicity and effectiveness features that make for a very intuitive and
pragmatic solution to incorporating progressive compression and access into existing
pipelines in a rather nondisruptive fashion.  We envision that many community databases
could incorporate our framework by simply adding an extra ``component dimension'' to data
variables using some agreed upon naming convention.  We see such application integrations
as important avenues for future work.

%% file: acknowledgments.tex
\acknowledgments{
This work was performed under the auspices of the U.S.\ Department of Energy by Lawrence
Livermore National Laboratory under Contract DE-AC52-07NA27344 and was supported by the
Office of Science, Office of Advanced Scientific Computing Research.
}